\documentclass[10pt,twocolumn,twoside]{IEEEtran}

\IEEEoverridecommandlockouts                              

\usepackage{graphics} 
\usepackage{epsfig} 
\usepackage{mathptmx} 
\usepackage{times} 
\usepackage{amsmath} 
\usepackage{amssymb}  
\usepackage{dsfont}
\usepackage{subfigure}
\usepackage{cancel}
\usepackage{margins}
\usepackage{mathtools}
\usepackage{multirow}  
\usepackage{setspace}
\newtheorem{thm}{Theorem}

\newtheorem{cor}{Corollary}
\newtheorem{lem}{Lemma}
\newtheorem{remark}{Remark}

\title{Collective Circular Motion of Multi-Agent Systems in Synchronized and Balanced Formations With Second-Order Rotational Dynamics}

\author{Anoop Jain and Debasish Ghose 
\thanks{A. Jain is a graduate student at the Guidance, Control and Decision System Laboratory (GCDSL) in the Department of
                 Aerospace Engineering, Indian Institute of Science,
                  Bangalore, India (email: anoopj@aero.iisc.ernet.in).}
\thanks{D. Ghose is a Professor at the Guidance, Control and Decision System Laboratory (GCDSL) in the Department of
              Aerospace Engineering, Indian Institute of
              Science, Bangalore, India (email: dghose@aero.iisc.ernet.in).}
\thanks {This work is partially supported by Asian Office of Aerospace Research and Development
(AOARD).}}

\begin{document}
\maketitle

\begin{abstract}
This paper considers the collective circular motion of multi-agent systems in which all the agents are required to traverse different circles or a common circle at a 
prescribed angular velocity. It is required to achieve these collective motions with the heading angles of the agents synchronized or balanced. In synchronization, the agents and their centroid have a common velocity direction, while in balancing, the movement of agents causes the location of the centroid to become stationary. The agents considered are initially moving at unit speed around individual circles at different angular velocities. It is assumed that the agents are subjected to limited communication constraints, and exchange relative information according to a time-invariant undirected graph. We present suitable feedback control laws for each of these motion coordination tasks by considering a second-order rotational dynamics of the agent. Simulations are given to illustrate the theoretical findings.
\end{abstract}



\section{Introduction}
There are various engineering applications such as tracking, surveillance, environmental monitoring, searching, sensing and data collection, where it is required for the multi-agent systems to perform a particular collective motion \cite{Beard2008}$-$\cite{Cortes2004}. A multi-agent system might comprise ground vehicles, air vehicles, underwater vehicles or a combination of these. In this article, we focus on achieving collective circular motion that can be applied in the scenario where vehicles are required to enclose, capture, secure or monitor a target or a search region.

Motivated by these applications, the collective motion where all the agents traverse $i)$ different circles, or $ii)$ a common circle at the prescribed angular velocity along with their heading angles in synchronized or in balanced states, are considered in this paper. Synchronization refers to the situation when all the agents, at all times, move in a common direction. A complementary notion of synchronization is balancing, in which all the agents move in such a way that their centroid, which is the average position of all the agents, remains fixed. It is evident in synchronized formation that agents and their centroid move in the same direction. Note that, in this paper, ``collective motion" and ``formation'' are used interchangeably.

Earlier work in \cite{Sepulchre2007} and \cite{Sepulchre2008} has focused on achieving synchronized and balanced formations in a group of agents under all-to-all and limited communication scenarios, respectively. In these papers, it is considered that the angular velocities of initial rotations of all the agents are the same and remains constant at all times. Recently, the effect of heterogeneity in various aspects have been studied in the literature. For example, \cite{Seyboth2014} considers nonidentical linear velocities of the agents, and \cite{AJain2015} considers heterogeneous control gains. In a similar spirit, in this paper, we consider that the angular velocities of the initial rotational motion of the agents are nonidentical and are allowed to vary with time. This more general scenario is addressed in this paper. In a similar context, the authors in \cite{Seth2013}, by assuming an all-to-all coupling among agents, propose feedback controls to stabilize synchronized and balanced circular formations at a desired angular velocity. However, unlike \cite{Seth2013}, in this paper, we further assume that the communication among agents is restricted and can be modeled as a time-invariant and undirected graph. Some related work, but with all-to-all communication, has been presented in \cite{Jain2015}.

There exists an ample literature related to the study of collective circular formation control. In \cite{Ceccarelli2008}, control laws are proposed to stabilize collective circular motion of nonholonomic vehicles around a virtual reference beacon, which is either stationary or moving. In \cite{Liu2016}, authors propose a distributed circular formation control law for ring-networked nonholonomic vehicles with local coordinate frames. In \cite{chen2011}, Chen and Zhang propose a decentralized control algorithm to form a class of collective circular motion, in which the vehicles are evenly distributed over the motion circle, and have the same rotational radius. The latter assumption is relaxed in \cite{chen2013}, where the agents move in circles around a common center, but with different radii.
In \cite{Arranz2009} and, \cite{Arranz2010}, the control algorithms to stabilize the collective motion around a circular orbit, which has either a fixed radius and time-varying center \cite{Arranz2009}, or a fixed center and time-varying radius \cite{Arranz2010}, are proposed. An extension of these results is given in \cite{Arranz2014}, where a new framework based on affine transformations is discussed to achieve more complex time-varying formations.
In \cite{Xu2013}, the splay circular formation, characterized by equally spaced arrangement of multiple robots, is stabilized by using a modified Kuramoto model \cite{Strogatz2000}. The stabilization of circular motion under cyclic pursuit is given in the seminal paper \cite{Marshall2004}, and also discussed in \cite{Ding2012} under dynamically adjustable control gains. Moreover, the cyclic pursuit problem of vehicles with heterogenous constant linear velocities is considered in \cite{Sinha2007}. In \cite{Summers2009}, a Lyapunov guidance vector field approach is used to guide a team of unmanned aircraft to fly a circular orbit around a moving target with prescribed inter-vehicle angular spacing. The circumnavigation problem for a team of unicycle-type agents, with the goal of achieving specific circular formations and circling on different orbits centered at a target of interest, is studied in \cite{Zhenga2015}.

It is to be noted that in the literature described above, most of the attention is towards achieving a particular type of collective circular formation. However, in the present work, the emphasis is given toward achieving the same along with a particular arrangement of the heading angles of the agents which could be a synchronized, balanced or a combination of both (usually called as symmetric phase pattern). These formations serves as the motion primitives, and can be utilized to get more general motion patterns \cite{Sepulchre2007}.

The main contribution of this paper is to propose a limited communication based control strategy to stabilize aforementioned collective circular motion of a group of agents with their phase arrangements either in synchronized or in balanced formation, while allowing the angular velocities of individual, initial circular motions, performed by the agents, to be different. With this purpose, in this paper, we consider identical agents moving in a planar space at constant unit linear speed with second-order rotational dynamics. Thus, the dynamics of each agent is represented by a state vector, which includes the position, heading angle and angular velocity of each agent as its elements. The second-order rotational model is particularly relevant in the context of planar rigid-body motion, where a dynamic vehicle model must account not only for motion of the agent's center of mass, but also for rotational motion about the center of mass \cite{Mellish2010}. We use second-order rotational model to derive feedback controls that are adequate to regulate the orientations as well as the angular velocities of the agents.

The outline of the paper is as follows. In Section II, we describe the system model and formulate the problem. In Section III, control laws are proposed to stabilize collective motion of agents on different circles at desired angular velocity with their phase arrangement either in synchronized or in balanced states. The control laws to stabilize collective motion around a common circle of desired radius as well as center with their phase arrangement, again either in synchronized or in balanced states, is proposed in Section IV. The control strategy to stabilize symmetric balanced patterns is proposed in Section V. In Section VI, we combine the results of the previous sections and propose control algorithms to stabilize symmetric circular formations suitable for mobile sensor network applications. Finally, Section VII concludes the paper.

\begin{figure}
\centering
\subfigure[]{\includegraphics[scale=1.38]{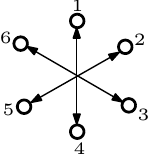}}\label{1a}\hspace{1.5cm}
\subfigure[]{\includegraphics[scale=1.38]{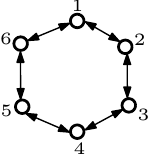}}
\caption{Undirected circulant graphs for $N = 6$. Both (a) and (b) are circulant graphs but only (b) is connected.}
\label{Analysis of gains}
\end{figure}

\section{System Description and Problem Statement}

\subsection{Agent Model}
Similar to \cite{Seth2013} and \cite{Mellish2010}, the collective second-order rotational dynamics of $N$ identical agents, moving in a planar space, each assumed to have unit mass and unit linear speed, is represented as
\begin{subequations}\label{modelNew}
\begin{align}
\label{modelNew1}\dot{r}_k &= e^{i\theta_k}\\
\label{modelNew2} \dot{\theta}_k & = \omega_k \\
\label{modelNew3}\dot{\omega}_k &= u_k, ~~~ k = 1, \ldots, N.
\end{align}
\end{subequations}
Here, complex notations are used to describe the position and velocity of each agent. For $k = 1, \ldots, N$, the position of the $k^\text{th}$ agent is $r_k \in \mathbb{C}$, while the velocity of the $k^\text{th}$ agent is $\dot{r}_k = e^{i\theta_k} = \cos\theta_k + i\sin\theta_k \in \mathbb{C}$, where, $\theta_k$ is the orientation of the (unit) velocity vector of the $k^\text{th}$ agent from the real axis, and $i = \sqrt{-1}$ represents the standard complex number. The orientation $\theta_k$, of the velocity vector, which is also referred to as the phase of the $k^\text{th}$ agent \cite{Strogatz2000}, represents a point on the unit circle $\mathbb{S}^1$.

In \eqref{modelNew}, $\omega_k \in \mathbb{R}$ is the angular velocity of the $k^\text{th}$ agent, which is determined by the feedback control $u_k \in \mathbb{R}$. If the control law $u_k$ is constant and equal to $\omega_k \neq 0$, then the $k^\text{th}$ agent travels at constant unit linear speed on a circle of radius $\rho_k = |\omega_k|^{-1}$. The direction of rotation around the circle is determined by the sign of $\omega_k$. If $\omega_k > 0$, then the $k^\text{th}$ agent rotates in the anticlockwise direction, whereas, if $\omega_k < 0$, then the $k^\text{th}$ agent rotates in the clockwise direction.

Let the initial motion of all the agents with dynamics \eqref{modelNew} be governed by the open-loop control $u_k = 0,~\forall k$. In this situation, the $k^\text{th}$ agent moves in a circular orbit of radius $|\omega_k|^{-1}$ with angular velocity $\omega_k$. Our aim is to seek a feedback control $u_k,~\forall k$ such that the collective motion of agents, subjected to limited communication constraints represented by a time-invariant undirected graph, converge to a circular motion at desired angular velocity (and hence desired radius of the circling orbit since radius of rotation = $|\text{angular velocity}|^{-1}$ for an agent circling at unit linear speed) with their phase angles either in synchronized or in balanced states. We assume that agents can exchange information only about their orientations according to the underlying interaction network, and they are globally provided the information about the desired angular velocity $\Omega_d$. In addition, when it is required for the agents to move around a common circle, the information about the desired center $c_d$ (of the common circle) is also globally provided to them. Note that issue of collision avoidance among agents is not considered in this work.

\subsection{Notations}
We introduce a few additional notations that are used in this paper. We use bold face letters $\pmb{r} = (r_1, \ldots, r_N)^T \in \mathbb{C}^N$, $\pmb{\theta} = (\theta_1, \ldots, \theta_N)^T \in \mathbb{T}^N$, where, $\mathbb{T}^N$ is the $N$-torus, which is equal to $\mathbb{S}^1 \times \ldots \times \mathbb{S}^1$ ($N$-times), and $\pmb{\omega} = (\omega_1, \ldots, \omega_N)^T \in \mathbb{R}^N$, to represent the vectors of length $N$ for the agents' positions, headings and angular velocities, respectively. Next, we define the inner product $\left<z_1, z_2\right>$ of two complex numbers $z_1, z_2 \in \mathbb{C}$ as $\left<z_1, z_2\right> = \textrm{Re}(\bar{z}_1z_2)$, where $\bar{z}_1$ represents the complex conjugate of $z_1$. For vectors, we use the analogous boldface notation $\left<\pmb{w}, \pmb{z}\right> = \textrm{Re}(\pmb{w^*}\pmb{z})$ for $\pmb{w}, \pmb{z} \in \mathbb{C}^N$, where $\pmb{w^*}$ denotes the conjugate transpose of $\pmb{w}$. The norm of $\pmb{z} \in \mathbb{C}^N$ is defined as $\|\pmb{z}\| = \left<\pmb{z}, \pmb{z}\right>^{1/2}$. The vectors $\pmb{0}$ and $\pmb{1}$ are used to represent by $\pmb{0} = (0,0,\ldots,0)^T \in \mathbb{R}^N$, and $\pmb{1} = (1,1,\ldots,1)^T \in \mathbb{R}^N$, respectively.


\subsection{Representation of Limited Communication Topology}
In the framework of multiagent systems, communication among agents is described by means of a graph. A graph is a pair $\mathcal{G} = (\mathcal{V}, \mathcal{E})$, where $\mathcal{V} = \{v_1, \ldots, v_N\}$ is a set of $N$ nodes or vertices and $\mathcal{E} \subseteq V\times V$ is a set of edges or links. Elements of $\mathcal{E}$ are denoted as $(v_j, v_k)$ which is termed an edge or a link from $v_j$ to $v_k$. An undirected link between nodes $v_j$ and $v_k$ indicates that the information can be shared from node $v_j$ to node $v_k$ and vice versa. A graph $\mathcal{G}$ is called an undirected graph if it consists of only undirected links. The node $v_j$ is called a neighbor of node $v_k$ if the link $(v_j, v_k)$ exists in the graph $\mathcal{G}$. In this article, the set of neighbors of node $v_j$ is represented by $\mathcal{N}_j$. A complete graph is an undirected graph in which every pair of nodes is connected, that is, $(v_j, v_k) \in \mathcal{E}$, $\forall j, k \in N$. The Laplacian of a graph $\mathcal{G}$, denoted by $\mathcal{L} = [l_{jk}] \in \mathbb{R}^{N\times N}$, is defined as \cite{Bai2011},
\[
    l_{jk}=
\begin{dcases}
    |\mathcal{N}_j|, & \text{if } j = k\\
    -1,              & \text{if } k \in \mathcal{N}_j\\
    0                & \text{otherwise}
\end{dcases}\label{laplacian}
\]
where, $|\mathcal{N}_j|$ is the cardinality of the set $\mathcal{N}_j$. This definition allows the representation of the several properties of a graph in the form of matrix properties of its Laplacian $\mathcal{L}$. It is well known that the Laplacian $\mathcal{L}$ of an undirected and connected graph $\mathcal{G}$ is $(\text{P}1)$ symmetric and positive semi-definite, and $(\text{P}2)$ has an eigenvalue of zero associated with the eigenvector $\pmb{1}$, that is, $\mathcal{L}\pmb{x} = 0$ iff $\pmb{x} = \pmb{1} x_0$.

In this article, we will also use the notion of a circulant graph. A graph $\mathcal{G}$ is circulant if and only if its Laplacian $\mathcal{L}$ is a circulant matrix, that is, $\mathcal{L}$ is completely defined by its first row \cite{Biggs1994}. Each subsequent row of a circulant matrix is the previous row shifted one position to the right with the first entry of the row equal to the last entry of the previous row. An example of an undirected circulant graph, consisting of $6$ nodes, is shown in Fig.~\ref{Analysis of gains}. Note that the Laplacian for the graphs in Figs.~1(a) and 1(b) are, respectively, given by $\mathcal{L}_a = \text{circ}(1, 0, 0, -1, 0, 0)$ and $\mathcal{L}_b = \text{circ}(2, -1, 0, 0, 0, -1)$, where, $\text{circ}(z)$ represents the circulant matrix with $z$ being its first row. As both $\mathcal{L}_a$ and $\mathcal{L}_b$ are circulant matrices, both the graphs shown in Fig.~\ref{Analysis of gains} are circulant, but only the graph shown in Fig.~1(b) is connected.

Now, we state the following lemma from \cite{Paley2007}, which describes various properties of an undirected circulant graph that will be useful in proving the results in this paper.

\begin{lem}\label{lemma1}
Let $\mathcal{L}$ be the Laplacian of an undirected circulant graph $\mathcal{G} = (\mathcal{V}, \mathcal{E})$ with $N$ vertices. Set $\phi_k = (k-1)2\pi/N$ for $k = 1, \ldots, N$. Then, the vectors
\begin{equation}
\pmb{f}^{(l)} = e^{i(l-1)\phi},~~l = 1, \ldots, N,
\end{equation}
define a basis of $N$ orthogonal eigenvectors of $\mathcal{L}$. The unitary matrix $F$ whose columns are the $N$ (normalized) eigenvectors $(1/\sqrt{N})\pmb{f}^{(l)}$ diagonalize $\mathcal{L}$, that is, $\mathcal{L} = F\Lambda F^\ast$, where $\Lambda = \text{diag}\{0, \lambda_2, \ldots,  \lambda_N\} \geq 0$ is the (real) diagonal matrix of the eigenvalues of $\mathcal{L}$.
\end{lem}


\section{Control Design}
The design of control laws is described in this section. At first, a phase potential $W_1(\pmb{\theta})$ is described, the minimization of which corresponds to synchronized formation, and its maximization corresponds to balanced formation. Then, a potential function $G(\pmb{\omega})$ whose minimization results in the collective motion of all the agents at a desired angular velocity, is proposed. Finally, the control law $u_k$ is obtained by minimizing a composite potential function consisting of $W_1(\pmb{\theta})$ and $G(\pmb{\omega})$ as described below.

\subsection{Achieving Synchronized and Balanced Formations}
The average linear momentum of a group of agents plays an important role in stabilizing their synchronized and balanced formations. It is maximized in synchronized formation and minimized in balanced formation. From \eqref{modelNew}, the average linear momentum, $p_\theta$, of a group of $N$-agents, is given by,
\begin{equation}
\label{momentum}p_\theta = \frac{1}{N}\sum_{k=1}^{N}e^{i\theta_k} = |p_\theta|e^{i\Psi},
\end{equation}
which is also referred to as the phase order parameter \cite{Strogatz2000}. The phase arrangement $\pmb{\theta}$ is synchronized if the modulus of the phase order parameter \eqref{momentum} equals one, that is, $|p_\theta| = 1$. The phase arrangement $\pmb{\theta}$ is balanced if the phase order parameter \eqref{momentum} equals zero, that is, $p_\theta = 0$ \cite{Strogatz2000}.

Thus, the stabilization of synchronized and balanced formations is accomplished by considering the potential
\begin{equation}
\label{all-to-all}U(\pmb{\theta}) = ({N}/{2})|p_\theta|^2,
\end{equation}
which reaches its unique minimum when $p_\theta = 0$ (balanced) and its unique maximum when all phases are identical (synchronized). Based on this potential function, the design of control law for the all-to-all communication among agents may be accomplished \cite{Jain2015}. However, in order to account for limited communication among agents, we modify the potential function \eqref{all-to-all} in the following manner \cite{Sepulchre2008}.

Let $P = I_N - (1/N)\pmb{1}\pmb{1}^T$, where, $I_N$ is an $N\times N$-identity matrix, be a projection matrix which satisfies $P^2 = P$. Let the vector $e^{i\pmb{\theta}}$ be represented by $e^{i\pmb{\theta}} = (e^{i\theta_1}, \ldots, e^{i\theta_N})^T \in \mathbb{C}^N$. Then,
\begin{equation}
\label{eq1}Pe^{i\pmb{\theta}} = \left(I_N - \frac{1}{N}\pmb{1}\pmb{1}^T\right)e^{i\pmb{\theta}} = e^{i\pmb{\theta}} - p_\theta\pmb{1}.
\end{equation}
Using \eqref{eq1}, one can obtain
\begin{equation}
\label{identity}\|P e^{i\pmb{\theta}}\|^2 = \left<e^{i\pmb{\theta}}, P e^{i\pmb{\theta}}\right> = N(1 - |p_\theta|^2),
\end{equation}
which is zero (minimum) when $|p_\theta| = 1$ (synchronized formation), and equates to $N$ (maximum) when $|p_\theta| = 0$ (balanced formation). Since $P$ is ($1/N$) times the Laplacian of the complete graph, the identity \eqref{identity} suggests that the optimization of $U(\pmb{\theta})$ in \eqref{all-to-all} may be replaced by the optimization of
\begin{equation}
\label{laplacian_potential}W_1(\pmb{\theta}) = Q_{\mathcal{L}}(e^{i\pmb{\theta}}) = \frac{1}{2}\left<e^{i\pmb{\theta}}, \mathcal{L}e^{i\pmb{\theta}}\right>,
\end{equation}
which is a Laplacian quadratic form associated with $\mathcal{L}$, and is positive semi-definite. Note that, for a connected graph, the quadratic form \eqref{laplacian_potential} vanishes only when $e^{i\pmb{\theta}} = e^{i\theta_c}\pmb{1}$, where $\theta_c \in \mathbb{S}^1$ is a constant (see property $\text{P}2$), that is, the potential $W_1(\pmb{\theta})$ is minimized in the synchronized formation.

The time derivative of $W_1(\pmb{\theta})$, along the dynamics \eqref{modelNew}, is
\begin{equation}
\label{W_Ldot}\dot{W}_1(\pmb{\theta}) = \sum_{k=1}^{N} \left(\frac{\partial W_1}{\partial \theta_k}\right) \dot{\theta}_k = \sum_{k=1}^{N} \left(\frac{\partial W_1}{\partial \theta_k}\right)\omega_k.
\end{equation}
Note that
\begin{align}
\nonumber\frac{\partial W_1}{\partial \theta_k} &= \frac{1}{2}\sum_{j=1}^{N}\frac{\partial}{\partial \theta_k}\left<e^{i\theta_j}, \mathcal{L}_je^{i\pmb{\theta}}\right>\\
\nonumber &= \frac{1}{2}\sum_{j=1}^{N}\left(\left<\frac{\partial e^{i\theta_j}}{\partial \theta_k}, \mathcal{L}_je^{i\pmb{\theta}}\right> + \left<e^{i\theta_j}, \frac{\partial (\mathcal{L}_j \nonumber e^{i\pmb{\theta}})}{\partial \theta_k}\right>\right)\\
\nonumber &= \frac{1}{2}\left(\left<ie^{i\theta_k}, \mathcal{L}_k e^{i\pmb{\theta}}\right> - \sum_{j \in \mathcal{N}_k} \left<e^{i\theta_j}, i e^{i\theta_k}\right>\right)\\
\label{note_that_2} &= \left<ie^{i\theta_k}, \mathcal{L}_k e^{i\pmb{\theta}}\right> = -\sum_{j \in \mathcal{N}_k} \sin(\theta_j - \theta_k),
\end{align}
where, $\mathcal{L}_k$ is the $k^\text{th}$ row of the Laplacian $\mathcal{L}$. Substituting \eqref{note_that_2} in \eqref{W_Ldot}, we get
\begin{equation}
\label{W_Ldot_new}\dot{W}_1(\pmb{\theta}) = \sum_{k=1}^{N} \left<ie^{i\theta_k}, \mathcal{L}_ke^{i\pmb{\theta}}\right>\omega_k.
\end{equation}

For the reasons which will be addressed in Section V, let us define the $m^\text{th}$ phase order parameter $p_{m\theta}$ and the phase potential $W_m(\pmb{\theta})$, respectively, as
\begin{equation}
\label{m_phase_order}p_{m\theta} = \frac{1}{mN}\sum_{k=1}^{N} e^{im\theta_k},
\end{equation}
\begin{equation}
\label{W_M}W_m(\pmb{\theta}) = \frac{1}{2}\left<e^{im\pmb{\theta}}, \mathcal{L}e^{im\pmb{\theta}}\right>,
\end{equation}
where, $m \in \mathbb{N} \triangleq \{1, 2, 3, \ldots\}$, and $e^{im\pmb{\theta}} = (e^{im\theta_1},\ldots, e^{im\theta_N})^T$.

Now, we state the following lemmas from \cite{Paley2007}, which describes various properties of the phase potential $W_m(\pmb{\theta})$ that will be useful in proving the results in this paper.

\begin{lem}\label{lemma2}
({\it Critical points of the Laplacian phase potential $W_m(\pmb{\theta})$}) Let $\mathcal{L}$ be the Laplacian of an undirected and connected graph $\mathcal{G} = (\mathcal{V}, \mathcal{E})$ with $N$ vertices. Consider the Laplacian phase potential $W_m(\pmb{\theta})$ defined in \eqref{W_M}. If $e^{im\pmb{\theta}}$ for all $m \in \mathbb{N}$ is an eigenvector of $W_m(\pmb{\theta})$, then $m\pmb{\theta}$ is a critical point of $W_m(\pmb{\theta})$, and $m\pmb{\theta}$ is either synchronized or balanced. The potential $W_m(\pmb{\theta})$ reaches its global minimum if and only if $m\pmb{\theta}$ is synchronized. If $\mathcal{G}$ is circulant, then $W_m(\pmb{\theta})$ reaches its global maximum in a balanced phase arrangement.
\end{lem}

\begin{IEEEproof}
The critical points of $W_m(\pmb{\theta})$ are given by the $N$ algebraic equations
\begin{equation}
\label{eq2}\frac{\partial W_m}{\partial \theta_k} = \left<ie^{im\theta_k}, \mathcal{L}_ke^{im\pmb{\theta}}\right> = 0,~~1\leq k \leq N.
\end{equation}
Let $e^{im\pmb{\theta}}$ be an eigenvector of $\mathcal{L}$ with eigenvalue $\lambda \in \mathbb{R}$. Then $\mathcal{L}e^{im\pmb{\theta}} = \lambda e^{im\pmb{\theta}}$, and
\begin{equation}
\label{critical_set_1}\frac{\partial W_m}{\partial \theta_k} = \left<ie^{im\theta_k}, \mathcal{L}_ke^{im\pmb{\theta}}\right> = \lambda\left<ie^{im\theta_k}, e^{im\theta_k}\right> = 0,~\forall k,
\end{equation}
which implies that $m\pmb{\theta}$ is a critical point of $W_m(\pmb{\theta})$. Since graph $\mathcal{G}$ is undirected, the Laplacian $\mathcal{L}$ is symmetric, and hence its eigenvectors associated with distinct eigenvalues are mutually orthogonal \cite{strang2007}. Since $\mathcal{G}$ is also connected, $\pmb{1}$ spans the kernel of $\mathcal{L}$. Therefore, the eigenvector associated with $\lambda = 0$ is $e^{im\pmb{\theta}} = e^{i\theta_c}\pmb{1}$ for any $\theta_c \in \mathbb{S}^1$, which implies $m\pmb{\theta}$ is synchronized. Also $\left<e^{im\pmb{\theta}}, \mathcal{L}e^{im\pmb{\theta}}\right> = 0$ if and only if $e^{im\pmb{\theta}} = e^{i\theta_c}\pmb{1}$ for any $\theta_c \in \mathbb{S}^1$. This proves the first part of the theorem.

Next, we assume that $\mathcal{G}$ is a circulant graph, then Lemma~\ref{lemma1} provides the following equivalent expression of the phase potential \eqref{W_M}.
\begin{equation}
W_m(\pmb{\theta}) = \frac{1}{2}\left<F^\ast e^{im\pmb{\theta}}, \Lambda F^\ast e^{im\pmb{\theta}}\right>.
\end{equation}
Let $\pmb{w} = F^\ast e^{im\pmb{\theta}}$, above equation can be written as
\begin{equation}
W_m(\pmb{\theta}) = \frac{1}{2}\left<\pmb{w}, \Lambda\pmb{w}\right> = \frac{1}{2}\sum_{k=2}^{N}|w_k|^2\lambda_k.
\end{equation}
Since matrix $F$ is unitary, $\|\pmb{w}\| = \|e^{im\pmb{\theta}}\| = \sqrt{N}$. Thus,
\begin{equation}
W_m(\pmb{\theta}) = \frac{1}{2}\left<\pmb{w}, \Lambda\pmb{w}\right> \leq \frac{N}{2} \lambda_{\text{max}},
\end{equation}
where, $\lambda_{\text{max}}$ is the maximum eigenvalue of $\mathcal{L}$. The maximum of $W_m(\pmb{\theta})$ is attained by selecting $e^{im\pmb{\theta}}$ as the eigenvector of $\mathcal{L}$ associated with the maximum eigenvalue. Since $e^{im\pmb{\theta}}$ is orthogonal to $\pmb{1}$, that is, it satisfies $\pmb{1}^Te^{im\pmb{\theta}} = 0$, and thus corresponds to the phase balancing of $m\pmb{\theta}$ (see \eqref{m_phase_order}). This completes the proof.
\end{IEEEproof}

\begin{remark}
For $m = 1$, the phase potential $W_1(\pmb{\theta})$ minimizes when all the phases synchronize. This state corresponds to the situation when all the agents, at all times, move in a common direction. On the other hand, the potential $W_1(\pmb{\theta})$ maximizes when all the phases balance. This state corresponds to the motion of all the agents about a fixed point such that $p_\theta = 0$. This fixed point is actually the position $R$ of the centroid of the agents since $\dot{R} = (1/N)\sum_{k=1}^{k} \dot{r}_k = p_\theta$. The use of the phase potential $W_m(\pmb{\theta})$ for $m > 1$ will be elaborated in Section V, where we address the stabilization of symmetric phase arrangements.
\end{remark}


\subsection{Achieving Desired Angular Velocity}
The agents, initially rotating at different angular velocities, are required to stabilize their collective motion at desired angular frequency $\Omega_d$ (and hence achieve the desired radius $\rho_d = |\Omega_d|^{-1}$). For this, we choose a potential function
\begin{equation}
G(\pmb{\omega}) = \frac{1}{2}\sum_{k=1}^{N}\left(\omega_k - \Omega_d\right)^2,
\end{equation}
which is minimized when $\omega_k = \Omega_d,~\forall k$.

The time derivative of $G(\pmb{\omega})$, along the dynamics \eqref{modelNew}, yields
\begin{equation}
\label{Gdot}\dot{G}(\pmb{\omega}) = \sum_{k=1}^{N}\left(\omega_k -
\Omega_d\right) \dot{\omega}_k = \sum_{k=1}^{N}\left(\omega_k -
\Omega_d\right) u_k
\end{equation}

\subsection{Composite Potential Function and Control Law}
In this subsection, the control law $u_k$, for the $k^\text{th}$ agent, is proposed by constructing a composite potential function, which ensures that all the agents travel around individual circles at a desired angular velocity $\Omega_d$ with their phases either in balanced or in synchronized states.

\begin{figure*}
\centerline{ \subfigure[]{\includegraphics[scale=0.3]{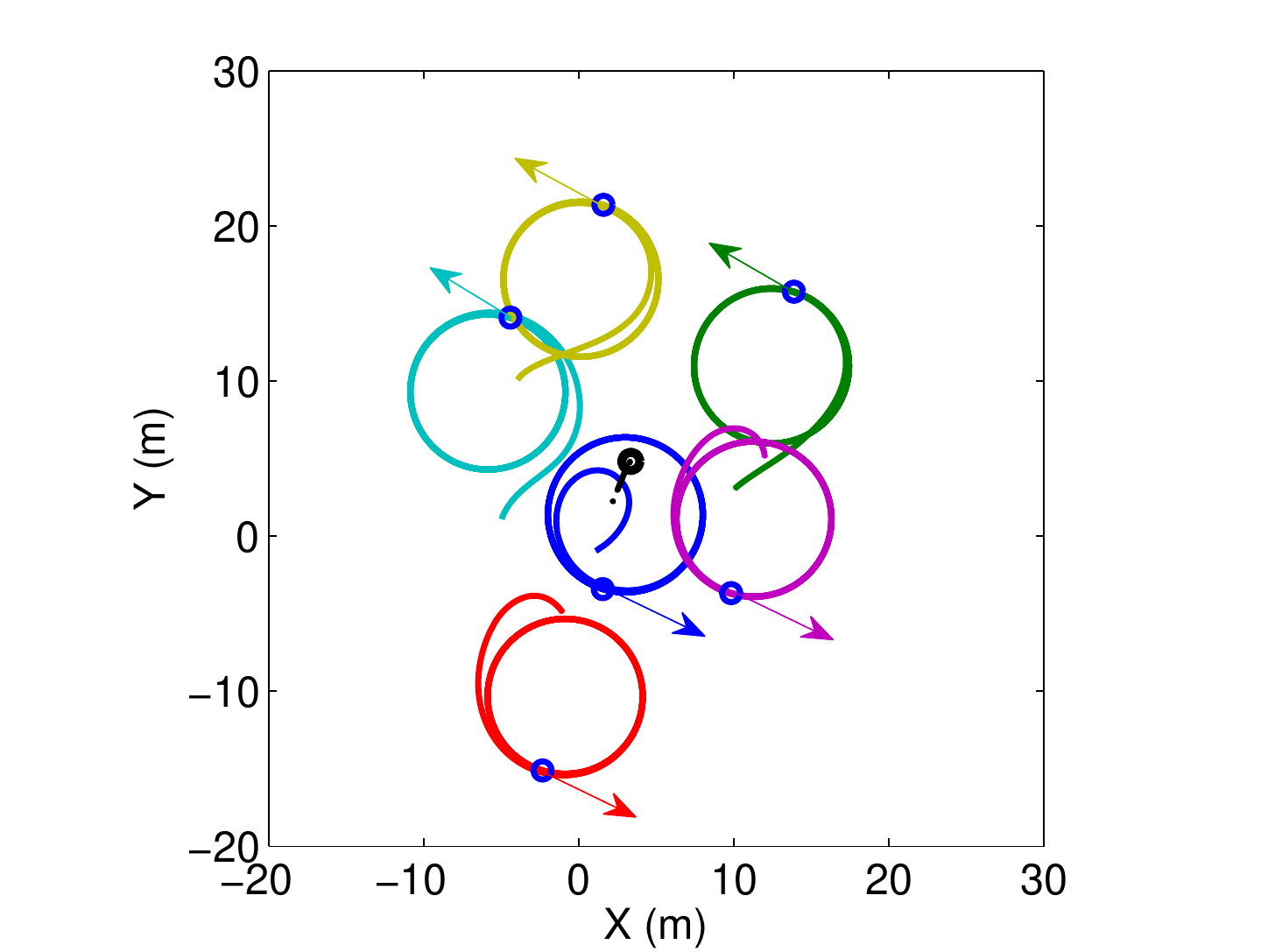}}\\
\subfigure[]{\includegraphics[scale=0.3]{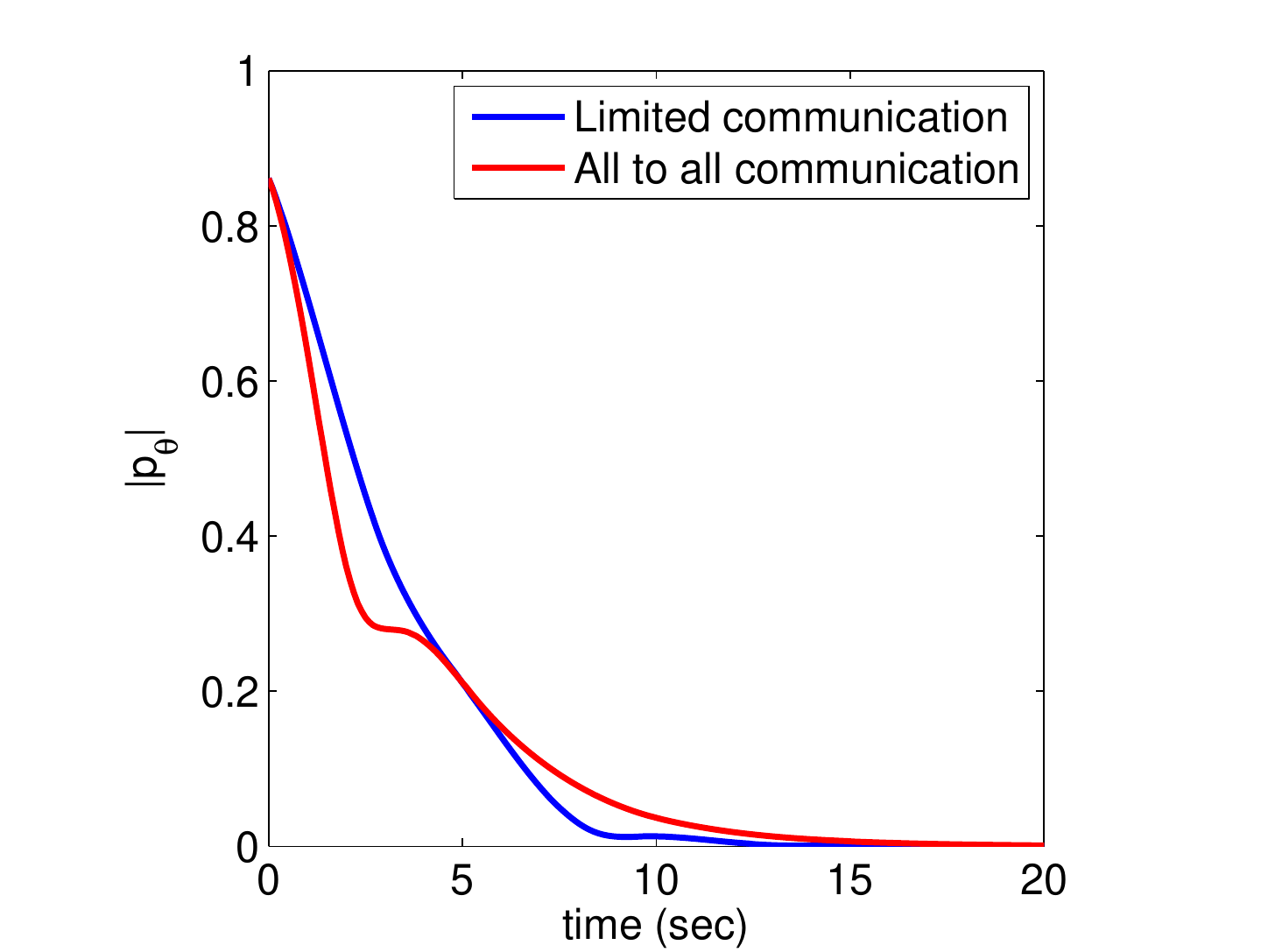}}\\
\subfigure[]{\includegraphics[scale=0.3]{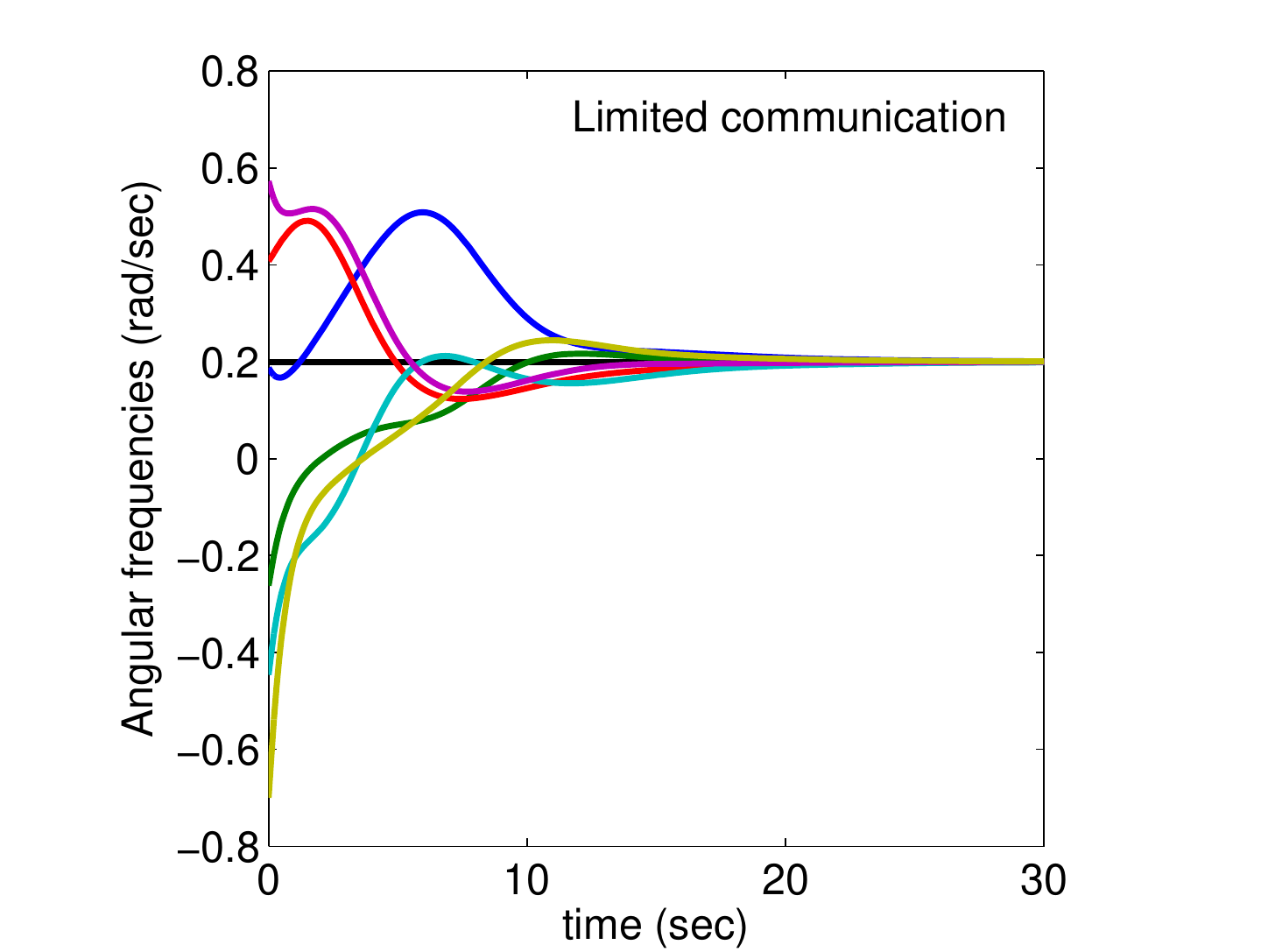}}\\
\subfigure[]{\includegraphics[scale=0.3]{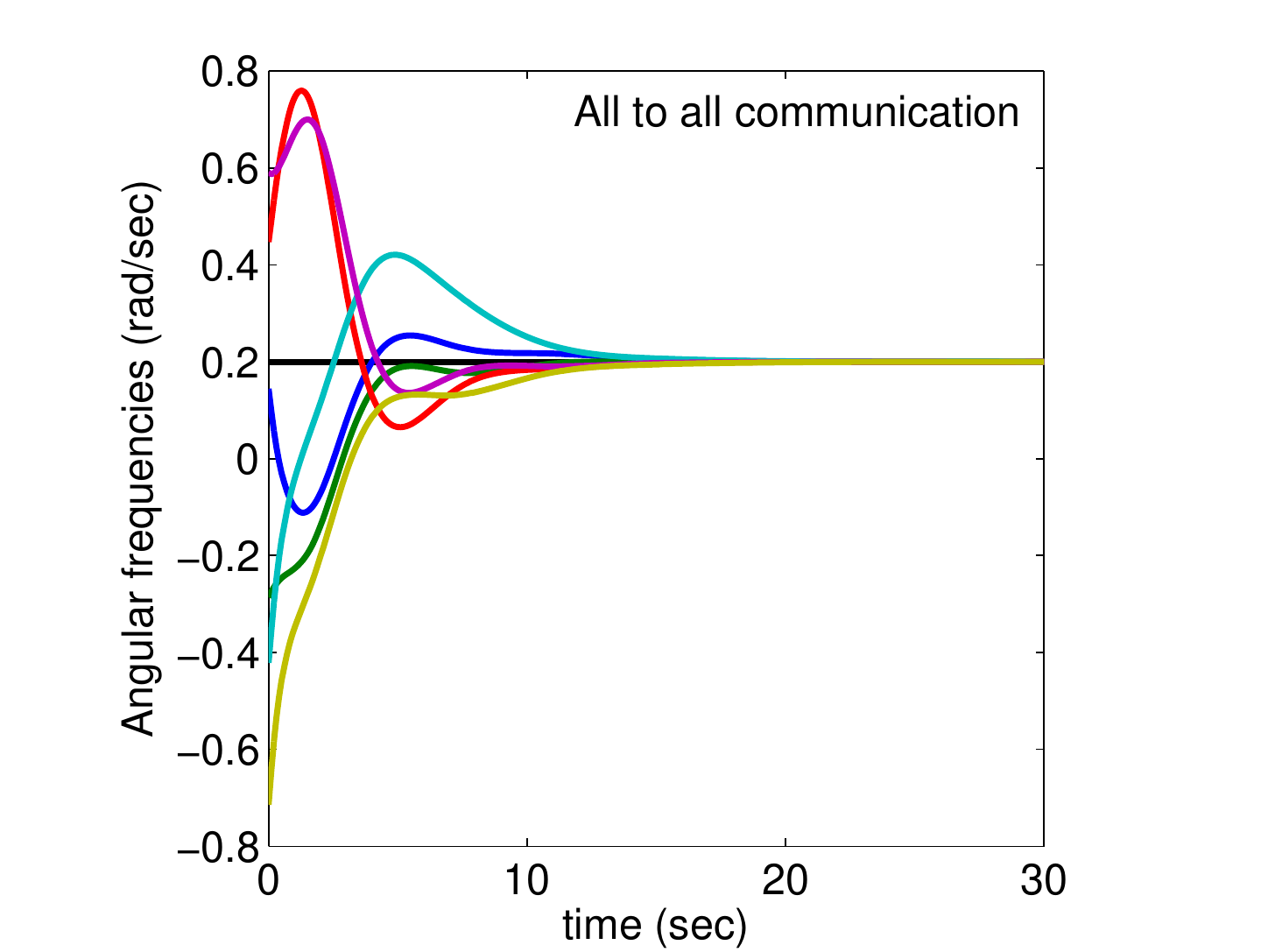}}}
\caption{Balancing of $N = 6$ agents, connected by a graph as shown in Fig.1(b), on individual circles, each having the desired radius $\rho_d = |\Omega_d|^{-1} = 5~ \text{m}$. $(a)$ Trajectories of the agents under the control laws \eqref{control1} with $K = 1$. $(b)$ Average linear momentum approaches zero with time. $(c)$ Consensus of angular speeds (frequencies) at desired value $\Omega_d = 0.2~ \text{rad/sec}$ (limited communication). $(d)$ Consensus of angular speeds at desired value $\Omega_d = 0.2~ \text{rad/sec}$ (all-to-all communication)}
\label{Balancing on different circles}
\end{figure*}
\begin{figure*}
\centerline{ \subfigure[]{\includegraphics[scale=0.3]{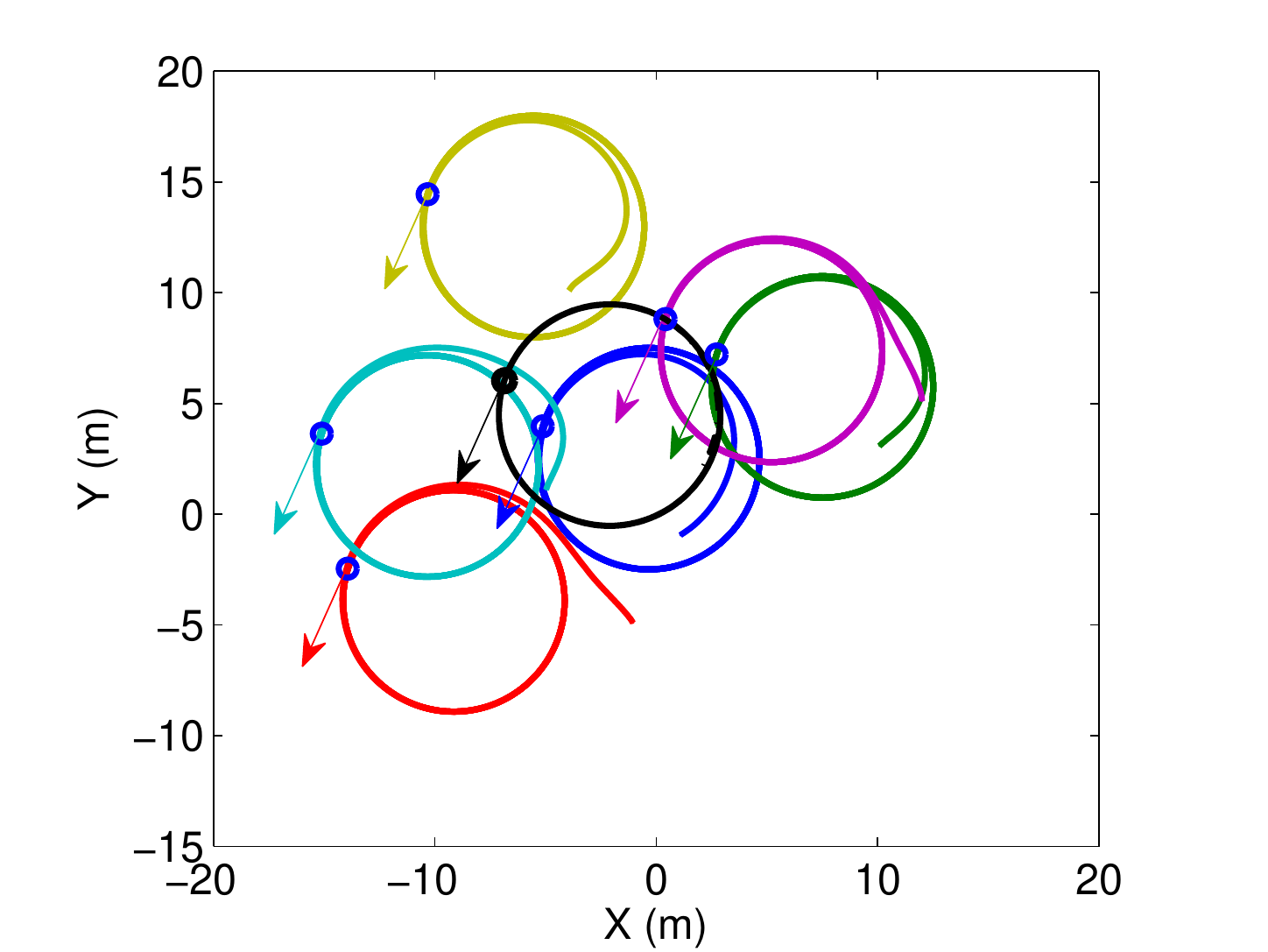}}\hspace{-0.6cm}
\subfigure[]{\includegraphics[scale=0.3]{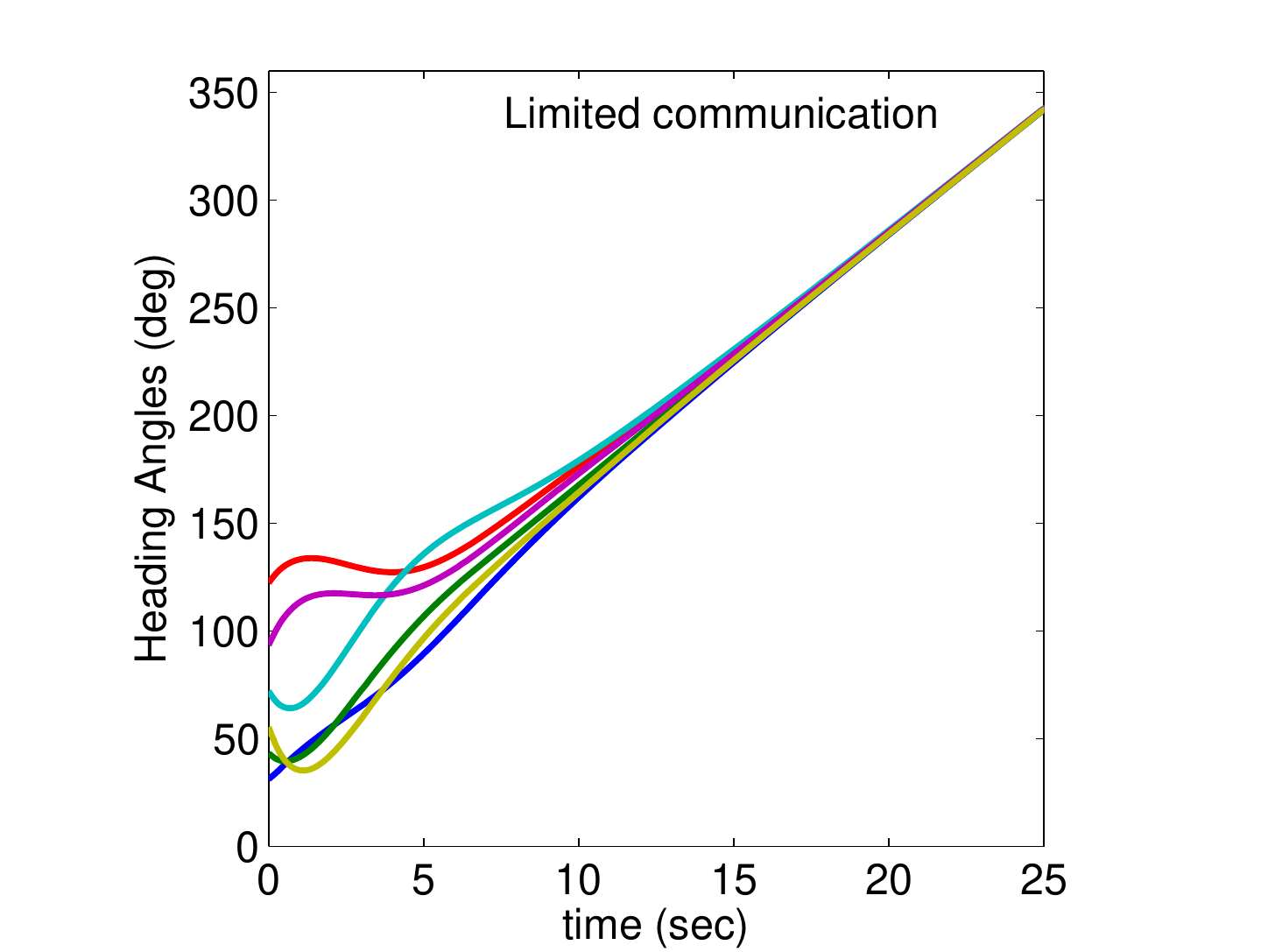}}\hspace{-0.8cm}
\subfigure[]{\includegraphics[scale=0.3]{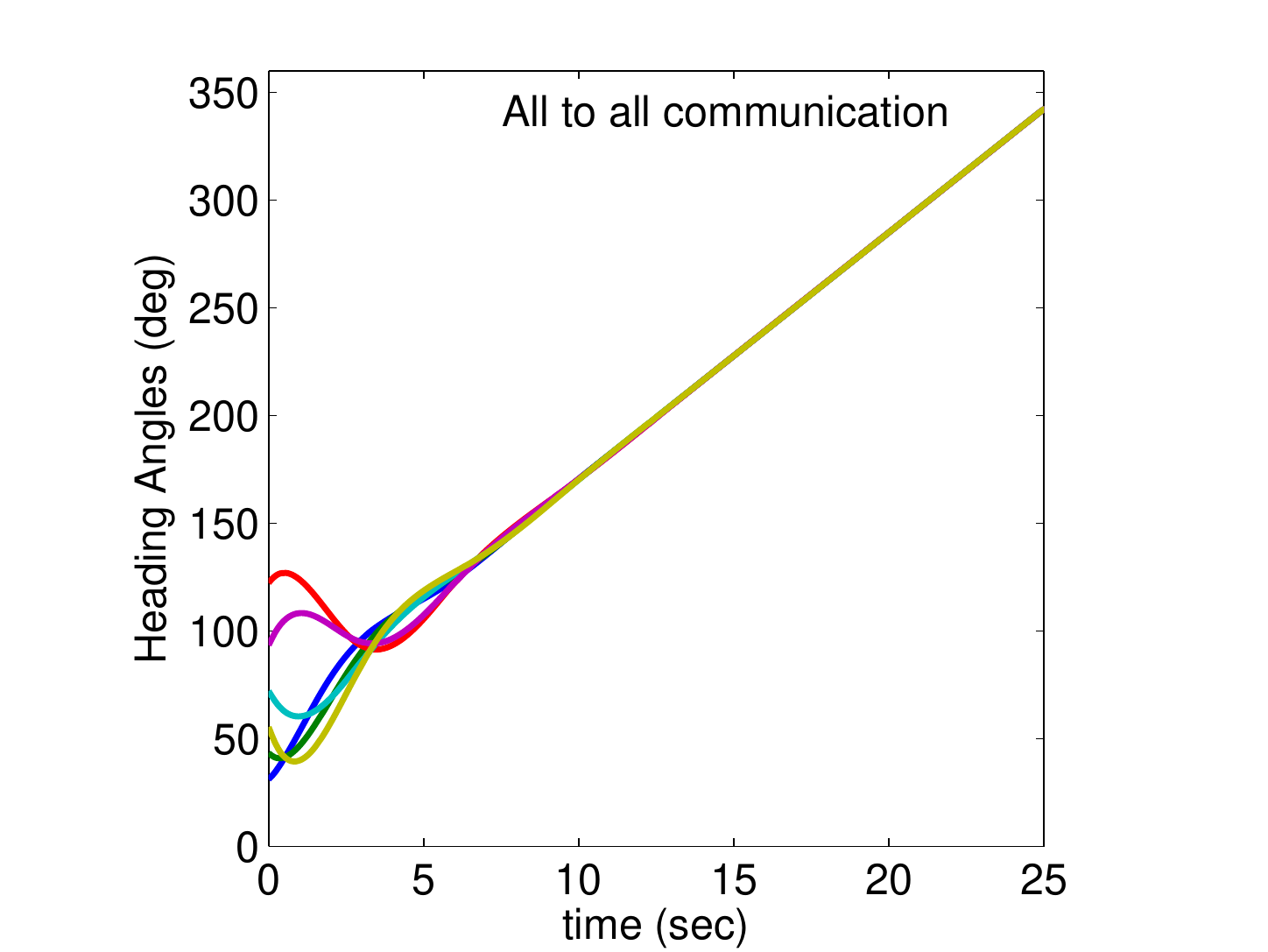}}\hspace{-0.8cm}
\subfigure[]{\includegraphics[scale=0.3]{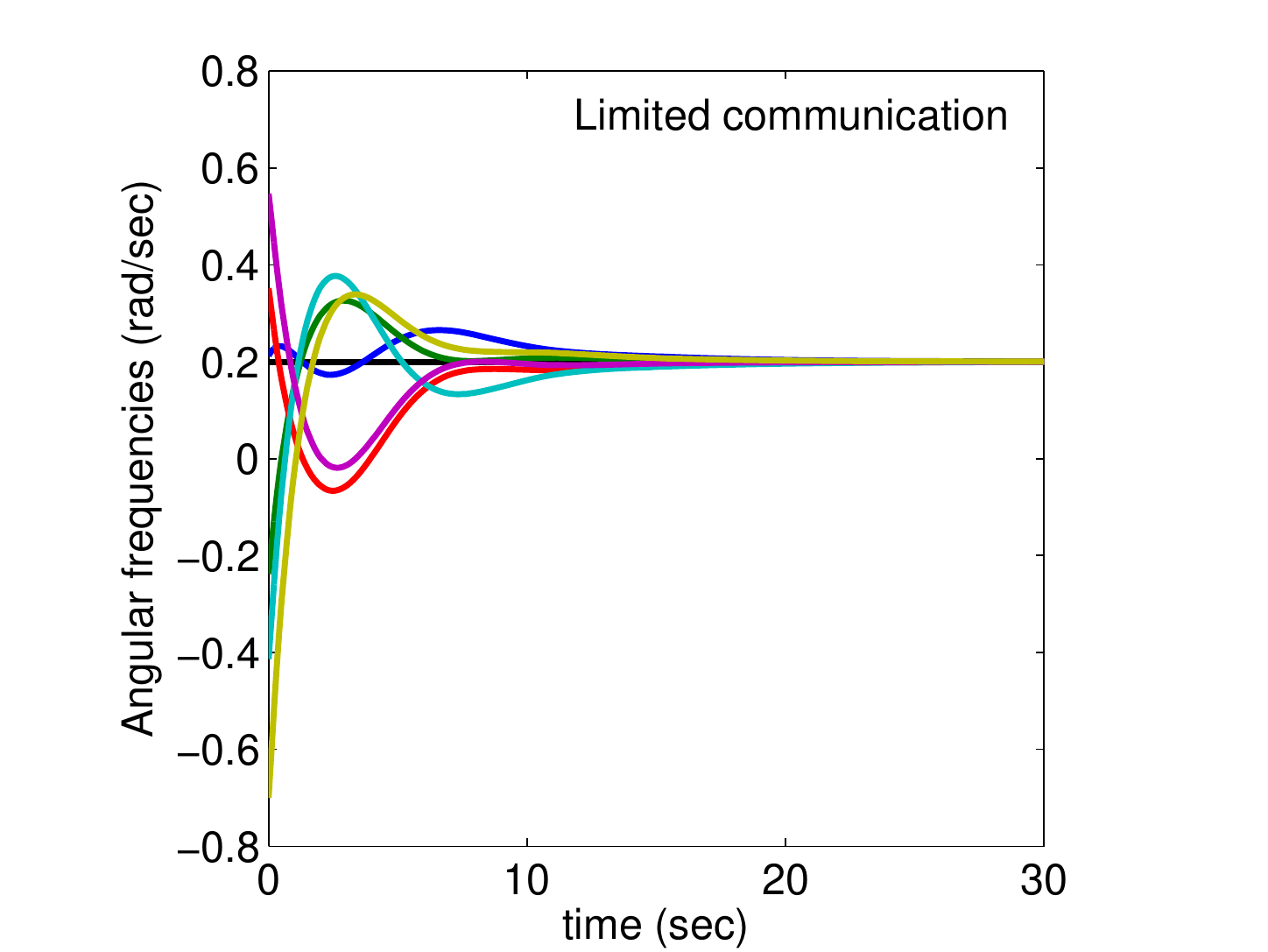}}\hspace{-0.8cm}
\subfigure[]{\includegraphics[scale=0.3]{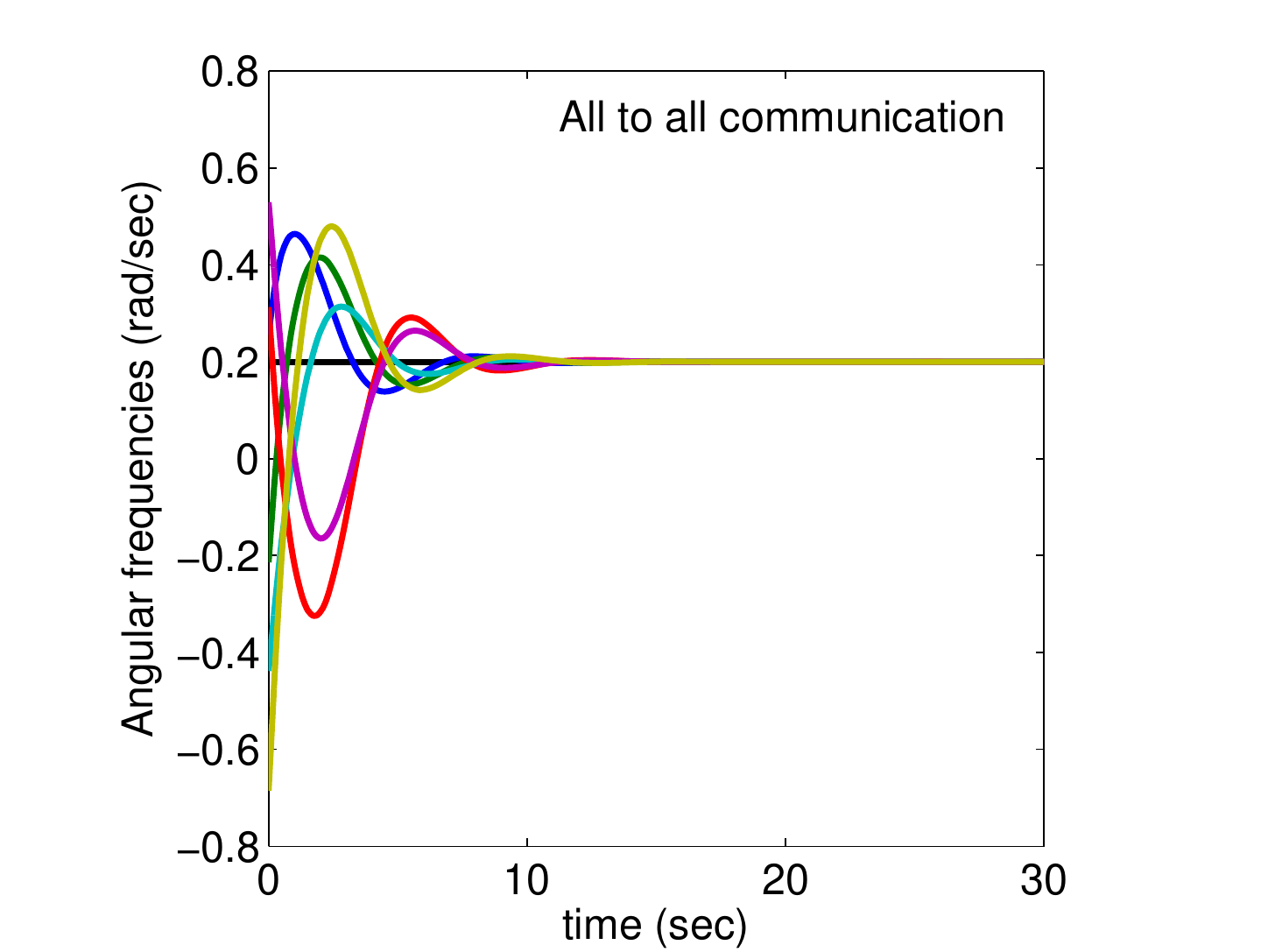}}}
\caption{Synchronization of $N = 6$ agents, connected by a graph as shown in Fig.1(b), on individual circles, each having the desired radius $\rho_d = |\Omega_d|^{-1} = 5~\text{m}$. $(a)$ Trajectories of the agents under the control laws \eqref{control2} with $K = -1$. $(b)$ Consensus in heading angles (limited communication). $(c)$ Consensus in heading angles (all-to-all communication). $(d)$ Consensus of angular speeds at desired value $\Omega_d = 0.2~ \text{rad/sec}$ (limited communication). $(e)$ Consensus of angular speeds at desired value $\Omega_d = 0.2~ \text{rad/sec}$ (all-to-all communication). Since the agents continue to rotate around their respective circles in synchronized fashion, their heading angles keep increasing with time.}
\label{Synchronization on different circles}
\end{figure*}

\begin{thm}\label{Theorem1}
Let $\mathcal{L}$ be the Laplacian of an undirected and connected circulant graph $\mathcal{G} = (\mathcal{V}, \mathcal{E})$ with $N$ vertices. Consider system dynamics \eqref{modelNew} with control law
\begin{equation}
\label{control1}u_k = -K\left((\omega_k - \Omega_d) - \frac{\partial W_1}{\partial \theta_k}\right),~\forall k.
\end{equation}
For $K>0$, all the agents converge to a collective motion in which they travel around individual circles of the same radius $\rho_d = |\Omega_d|^{-1}$ with their phase angles in balanced state.
\end{thm}

\begin{IEEEproof}
Consider a composite potential function
\begin{equation}
V_1(\pmb{\theta}, \pmb{\omega}) = K\left(\frac{N}{2}\lambda_\text{max} - W_1(\pmb{\theta})\right) + G(\pmb{\omega}); ~~K > 0.
\end{equation}
Note that $0 \leq W_1(\theta) \leq ({N}/{2})\lambda_\text{max}$ (Lemma~\ref{lemma2} for $m=1$) which ensures that $V_1(\pmb{\theta}, \pmb{\omega}) \geq 0$. Using \eqref{W_Ldot} and \eqref{Gdot}, the time derivative of the potential function $V_1(\pmb{\theta}, \pmb{\omega})$ along the dynamics \eqref{modelNew}, is
\begin{equation}
\dot{V}_1(\pmb{\theta}, \pmb{\omega}) = -K\sum_{k=1}^{N} \left(\frac{\partial W_1}{\partial \theta_k}\right) \omega_k + \sum_{k=1}^{N}\left(\omega_k -
\Omega_d\right) u_k.
\end{equation}
With the control law \eqref{control1}, the time derivative of $V_1(\pmb{\theta}, \pmb{\omega})$ results in
\begin{equation}
\label{V_dot_old}\dot{V}_1(\pmb{\theta}, \pmb{\omega}) = - K \sum_{k=1}^{N}(\omega_k - \Omega_d)^2 - K\Omega_d\sum_{k=1}^{N}\frac{\partial W_1}{\partial \theta_k}
\end{equation}
From \eqref{W_Ldot_new}, we note that
\begin{equation}
\label{relation}\sum_{k=1}^{N}\frac{\partial W_1}{\partial \theta_k} = - \sum_{k=1}^{N}\sum_{j \in \mathcal{N}_k} \sin(\theta_j - \theta_k) = 0.
\end{equation}
Using \eqref{relation}, \eqref{V_dot_old} becomes
\begin{equation}
\dot{V}_1(\pmb{\theta}, \pmb{\omega}) = - K \sum_{k=1}^{N}(\omega_k - \Omega_d)^2 \leq 0.
\end{equation}
Since $(\pmb{\theta}, \pmb{\omega}) \in \mathbb{T}^N \times \mathbb{R}^N$ is compact, it follows from the LaSalle's invariance theorem \cite{Khalil2000} that, for $K>0$, all the solutions of \eqref{modelNew} with the control law \eqref{control1} converge to the largest invariant set contained in $\{\dot{V}_1(\pmb{\theta}, \pmb{\omega}) = 0\}$, that is, the set
\begin{equation}
\label{invariant_set_b} \Delta_b = \left\{(\pmb{\theta}, \pmb{\omega}) ~|~ \omega_k = \Omega_d,~\forall k\right\}.
\end{equation}
In $\Delta_b$, $\omega_k = \Omega_d,~\forall k$, which implies that each agent rotates with angular speed $\Omega_d$. Moreover, since $u_k = \dot{\omega}_k = 0,~\forall k$ in the set $\Delta_b$, it implies from \eqref{control1} that $\left(\partial W_1 / \partial \theta_k\right) = 0,~\forall k$, which defines the critical points of $W_1(\pmb{\theta})$. The set $\Delta_b$ is itself the largest invariant set since
\begin{align}
\nonumber \frac{d}{dt} \left(\frac{\partial W_1}{\partial \theta_k}\right) &= \frac{d}{dt} \left<ie^{i\theta_k}, L_ke^{i\pmb{\theta}}\right>\\
\label{relation1}&= \left<-\Omega_de^{i\theta_k}, L_ke^{i\pmb{\theta}}\right> + \left<ie^{i\theta_k}, -i\Omega_dL_ke^{i\pmb{\theta}}\right> = 0,~\forall k,
\end{align}
on this set. Now, we analyze the critical points of $W_1(\pmb{\theta})$.

{\it Analysis of the critical points of $W_1(\pmb{\theta})$}: The critical points of $W_1(\pmb{\theta})$ are given by the $N$ algebraic equations
\begin{equation}
\label{eq3}\frac{\partial W_1}{\partial \theta_k} = \left<ie^{i\theta_k}, \mathcal{L}_ke^{i\pmb{\theta}}\right> = 0,~~1\leq k\leq N.
\end{equation}
For $m=1$, since \eqref{eq3} is the same as \eqref{eq2}, it follows from Lemma~\ref{lemma2} that $W_1(\pmb{\theta})$ minimizes in the synchronized state and maximizes in balanced state. Moreover, since maximization of the potential $W_1(\pmb{\theta})$ corresponds to the global minimum of $V_1(\pmb{\theta}, \pmb{\omega})$, balanced formation is asymptotically stable. This completes the proof.
\end{IEEEproof}


\begin{thm}\label{Theorem2}
Let $\mathcal{L}$ be the Laplacian of an undirected and connected circulant graph $\mathcal{G} = (\mathcal{V}, \mathcal{E})$ with $N$ vertices. Consider system dynamics \eqref{modelNew} with control law
\begin{equation}
\label{control2}u_k = K\left((\omega_k - \Omega_d) + \frac{\partial W_1}{\partial \theta_k}\right),~\forall k.
\end{equation}
For $K < 0$, all the agents converge to a collective motion in which they travel around individual circles of the same radius $\rho_d = |\Omega_d|^{-1}$ with their phase angles in synchronized state.
\end{thm}

\begin{IEEEproof}
Consider a composite potential function
\begin{equation}
V_2(\pmb{\theta}, \pmb{\omega}) = -K W_1(\pmb{\theta}) + G(\pmb{\omega}); ~~K < 0
\end{equation}
Using \eqref{W_Ldot_new} and \eqref{Gdot}, the time derivative of the potential function $V_2$ along the dynamics \eqref{modelNew}, is
\begin{equation}
\dot{V}_2(\pmb{\theta}, \pmb{\omega}) = -K\sum_{k=1}^{N} \left(\frac{\partial W_1}{\partial \theta_k}\right) \omega_k + \sum_{k=1}^{N}\left(\omega_k -
\Omega_d\right) u_k
\end{equation}
With the control law \eqref{control2}, the time derivative of $V_2$ results in
\begin{equation}
\label{v2_dot_old}\dot{V}_2(\pmb{\theta}, \pmb{\omega}) =  K \sum_{k=1}^{N}(\omega_k - \Omega_d)^2 - K\Omega_d\sum_{k=1}^{N}\frac{\partial W_1}{\partial \theta_k}.
\end{equation}
Using \eqref{relation}, \eqref{v2_dot_old} becomes
\begin{equation}
\dot{V}_2(\pmb{\theta}, \pmb{\omega}) =  K \sum_{k=1}^{N}(\omega_k - \Omega_d)^2 \leq 0.
\end{equation}
Since $(\pmb{\theta}, \pmb{\omega}) \in \mathbb{T}^N \times \mathbb{R}^N$ is compact, it follows from the LaSalle's invariance theorem \cite{Khalil2000} that, for $K<0$, all the solutions of \eqref{modelNew} with the control law \eqref{control2} converge to the largest invariant set contained in $\{\dot{V}_2(\pmb{\theta}, \pmb{\omega}) = 0\}$, that is, the set
\begin{equation}
\label{invariant_set_s} \Delta_s = \left\{(\pmb{\theta}, \pmb{\omega}) ~|~ \omega_k = \Omega_d,~\forall k\right\}.
\end{equation}
In $\Delta_s$, $\omega_k = \Omega_d,~\forall k$, which implies that each agent rotates with angular speed $\Omega_d$. Moreover, since $u_k = \dot{\omega}_k = 0,~\forall k$ in the set $\Delta_s$, it implies from \eqref{control1} that $\left(\partial W_1 / \partial \theta_k\right) = 0,~\forall k$, which defines the critical points of $W_1(\pmb{\theta})$. The set $\Delta_s$ is itself the largest invariant set since \eqref{relation1} holds. Following Lemma~\ref{lemma2}, since minimization of the potential $W_1(\pmb{\theta})$ corresponds to the global minimum of $V_2(\pmb{\theta}, \pmb{\omega})$, balanced formation is asymptotically stable. This completes the proof.
\end{IEEEproof}

\begin{remark}
In the previous analysis, the invariant sets $\Delta_b$ and $\Delta_s$, defined by \eqref{invariant_set_b} and \eqref{invariant_set_s}, respectively, look similar, however, they are different since $\Delta_b$ contains phases $\pmb{\theta}$ corresponding to the balanced formation, while $\Delta_s$ contains phases $\pmb{\theta}$ corresponding to the synchronized formation.
\end{remark}

{\it Example~1:} In this example, Theorems~\ref{Theorem1} and \ref{Theorem2} are demonstrated through simulation of $N=6$ agents connected via a graph as shown in Fig.~1(b). Let the initial positions, initial heading angles and nonidentical initial angular velocities of the agents be $\pmb{r}(0) = ((1, -1), (10, 3), (-1, -5), (-5, 1), (12, 5), (-4, 10))^T$, $\pmb{\theta}(0) = (30^\circ, 45^\circ, 120^\circ, 75^\circ, 90^\circ, 60^\circ)^T$, and $\pmb{\omega}(0) = (0.2, -0.3, 0.4, -0.5, 0.6, -0.8)^T$, respectively. Although the initial locations of the agents are given for representing the trajectories of the agents in the simulation, the locations themselves are not important so far as the objective of synchronization is concerned. Even with different locations, the convergence properties will be the same, although the trajectories will be different.

Fig.~\ref{Balancing on different circles} shows balancing of the agents around individual circles at desired angular velocity $\omega_d = 0.2~ \text{rad/sec}$, and hence desired radius $\rho_d = |\Omega_d|^{-1} = 5~ \text{m}$ of circular orbits as expected. On the other hand, Fig.~\ref{Synchronization on different circles} shows synchronization of agents on individual circles at the same desired angular frequency velocity. In all figures in this paper, the trajectory of the centroid is shown in black. Note that, in these figures, the convergence of the angular speeds (frequencies) to the desired value $\Omega_d$ is faster under all-to-all communication scenario as expected.


\section{Motion on a Common Circle}
\begin{figure}
\centering
\includegraphics[scale=0.7]{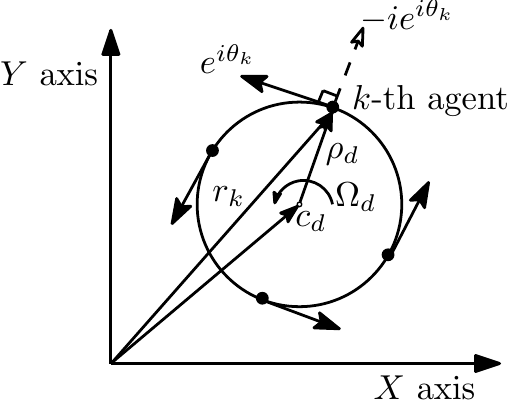}
\caption{Orientation of the $k^{th}$ agent on a circle of desired radius $\rho_d = {\Omega}^{-1}_d > 0$ and center $c_d$.}
\label{Orientation of the $k$-th agent on the desired circle}
\end{figure}

In this section, we achieve synchronization and balancing of a group of agents around a common circle of desired radius $\rho_d = |\Omega^{-1}_d|$ as well as center $c_d$, which is fixed. This situation is shown in Fig.~\ref{Orientation of the $k$-th agent on the desired circle}, where, the $k^\text{th}$ agent rotates in the anticlockwise direction on a circle of radius $\rho_d = |{\Omega}^{-1}_d|$ and center $c_d$. Without loss of generality, we assume $\Omega_d > 0$. Therefore, at equilibrium, all the agents, which may initially be rotating in clockwise and anticlockwise directions, move in the anticlockwise direction on a common circle of desired radius as well as center. The position $r_k$ of the $k$-th agent in Fig.~\ref{Orientation of the $k$-th agent on the desired circle}, is given by
\begin{equation}
\label{position}r_k = c_d - i \rho_d e^{i\theta_k}.
\end{equation}
In order to stabilize collective motion of all the agents around a common circle of radius $\rho_d = \Omega^{-1}_d > 0$ and center $c_d$, which is fixed, we introduce an error variable $e_k = r_k - (c_d - i \rho_d e^{i\theta_k}), \forall~k$, and choose a potential function as,
\begin{eqnarray}
\nonumber S(\pmb{r},\pmb{\theta}) &=& \frac{1}{2}\sum_{k=1}^{N} |e_k|^2 =
\frac{1}{2}\sum_{k=1}^{N}|r_k-c_d + i\rho_de^{i\theta_k}|^2\\
&=& \frac{1}{2}\sum_{k=1}^{N} \left<r_k-c_d + i\rho_de^{i\theta_k}, r_k-c_d + i\rho_de^{i\theta_k}\right>\label{S1},
\end{eqnarray}
which is non-negative and becomes zero whenever $r_k = c_d - i\rho_de^{i\theta_k}, \forall k$. It means that the minimization of  $S(\pmb{r},\pmb{\theta})$ corresponds to the situation when all the agents move on a common circle of radius $\rho_d = \Omega^{-1}_d > 0$ centered at the fixed point $c_d$.

The time derivative of the potential function \eqref{S1} along the dynamics \eqref{modelNew}, yields
\begin{equation}
\dot{S}(\pmb{r},\pmb{\theta}) = \sum_{k=1}^{N}\left<r_k-c_d-i\rho_de^{i\theta_k}, e^{i\theta_k}\left(1 - \rho_d\omega_k \right)\right>\label{Sdot1}
\end{equation}

Using linearity of inner product \cite{Rorres2011} in \eqref{Sdot1}, we get
\begin{eqnarray}
\nonumber\dot{S}(\pmb{r},\pmb{\theta}) &=& \sum_{k=1}^{N}\left<r_k-c_d, e^{i\theta_k}\right>\left(1 - \rho_d\omega_k\right) \\
&& -\sum_{k=1}^{N}\rho_d \left<ie^{i\theta_k}, e^{i\theta_k}\right>\left(1 - \rho_d\omega_k \right)\label{Sdot2}
\end{eqnarray}

Since $\left<ie^{i\theta_k}, e^{i\theta_k}\right> = 0$, \eqref{Sdot2} is simplified to
\begin{equation}
\label{Sdot}\dot{S}(\pmb{r},\pmb{\theta}) = \sum_{k=1}^{N}\left<r_k-c_d,
e^{i\theta_k}\right>\left(1 - \rho_d\omega_k \right).
\end{equation}

\begin{thm}\label{Theorem3}
Let $\mathcal{L}$ be the Laplacian of an undirected and connected circulant graph $\mathcal{G} = (\mathcal{V}, \mathcal{E})$ with $N$ vertices. Consider system dynamics \eqref{modelNew} with control law
\begin{equation}
\label{control3}u_k = -\kappa \rho_d(\omega_k - \Omega_d) + {\Omega}^2_d \left(\kappa \left<r_k-c_d, e^{i\theta_k}\right> + K \frac{\partial W_1}{\partial \theta_k}\right).
\end{equation}
For $K>0$ and $\kappa > 0$, all the agents converge to a circular formation in which they travel around a common circle of radius $\rho_d = \Omega_d^{-1} > 0$ and center $c_d$ in the anticlockwise direction with their phase angles in balanced state.
\end{thm}

\begin{IEEEproof}
Consider a composite potential function
\begin{equation}
U_1(\pmb{r}, \pmb{\theta}, \pmb{\omega}) = \kappa S(\pmb{r},\pmb{\theta}) + \rho_d K \left(\frac{N}{2}\lambda_\text{max} - W_1(\pmb{\theta})\right) + {\rho}^3_d G(\pmb{\omega});~~K>0, \kappa>0.\label{U_1}
\end{equation}
Using \eqref{W_Ldot_new}, \eqref{Gdot} and \eqref{Sdot}, the time derivative of the potential function $U_1(\pmb{r}, \pmb{\theta}, \pmb{\omega})$ along the trajectories of \eqref{modelNew}, is
\begin{align}
\nonumber & \dot{U}_1(\pmb{r}, \pmb{\theta}, \pmb{\omega}) = \kappa
\sum_{k=1}^{N}\left<r_k-c_d, e^{i\theta_k}\right>\left(1 -
\rho_d\omega_k\right) \\
& + K \sum_{k=1}^{N}\left(\frac{\partial W_1}{\partial \theta_k}\right) (-\rho_d\omega_k) + {\rho}^3_d \sum_{k=1}^{N}\left(\omega_k - \Omega_d\right) u_k\label{V1dot1}
\end{align}
Using \eqref{relation}, \eqref{V1dot1} can be rewritten as
\begin{align}
\nonumber\dot{U}_1(\pmb{r}, \pmb{\theta}, \pmb{\omega}) &= \sum_{k=1}^{N} \left(\kappa \left<r_k-c_d, e^{i\theta_k}\right> + K \sum_{k=1}^{N} \frac{\partial W_1}{\partial \theta_k}\right)\\
& \times \left(1 - \rho_d\omega_k \right) + {\rho}^3_d \sum_{k=1}^{N}\left(\omega_k - \Omega_d\right) u_k
\end{align}
Under control law \eqref{control3}, the time derivative of $U_1(\pmb{r}, \pmb{\theta}, \pmb{\omega})$ results in
\begin{equation}
\dot{U}_1(\pmb{r}, \pmb{\theta}, \pmb{\omega}) = - \kappa {\rho}^4_d \sum_{k=1}^{N} (\omega_k - \Omega_d)^2 \leq 0.
\end{equation}
Since $(\pmb{r}, \pmb{\theta}, \pmb{\omega}) \in \mathbb{R}^N \times \mathbb{T}^N \times \mathbb{R}^N$ is compact, it follows from the LaSalle's invariance theorem \cite{Khalil2000} that, for $K>0$ and $\kappa > 0$, all the solutions of \eqref{modelNew} with the control law \eqref{control3} converge to the largest invariant set contained in $\{\dot{U}_1(\pmb{r}, \pmb{\theta}, \pmb{\omega}) = 0\}$, that is, the set
\begin{equation}
\label{invariant_set_circular} \Gamma = \left\{(\pmb{r}, \pmb{\theta}, \pmb{\omega}) ~|~ \omega_k = \Omega_d,~\forall k\right\}.
\end{equation}
In $\Gamma$, $\omega_k = \Omega_d,~\forall k$, which implies that each agent rotates with angular speed $\Omega_d$. Moreover, since $u_k = \dot{\omega}_k = 0,~\forall k$, in the set $\Gamma$, it implies from \eqref{control3} that
\begin{equation}
\label{relation_control}\kappa \left<r_k-c_d, e^{i\theta_k}\right> + K\frac{\partial W_1}{\partial \theta_k} = 0,
\end{equation}
for all $k$. Let $\Gamma_b$ be the largest invariant set in $\Gamma$. Taking the time-derivative of \eqref{relation_control} in the set $\Gamma$ yields
\begin{equation}
\label{derivative}\kappa \frac{d}{dt} \left<r_k-c_d, e^{i\theta_k}\right> + K \frac{d}{dt} \left(\frac{\partial W_1}{\partial \theta_k}\right) = 0.
\end{equation}
Using \eqref{relation1}, \eqref{derivative} becomes
\begin{equation}
\label{derivative_1} \frac{d}{dt} \left<r_k-c_d, e^{i\theta_k}\right> = \left<r_k-c_d, i\Omega_de^{i\theta_k}\right> + \left<e^{i\theta_k}, e^{i\theta_k}\right> = 0.
\end{equation}
Since $\left<e^{i\theta_k}, e^{i\theta_k}\right> = 1$, \eqref{derivative_1} can be written as
\begin{equation}
\label{derivative_2} \left<r_k-c_d, i\Omega_d e^{i\theta_k}\right> = -1.
\end{equation}
It is easy to check that \eqref{derivative_2} is satisfied only if
\begin{equation}
\label{position_new}r_k = c_d - i\rho_d e^{i\theta_k},~\forall k,
\end{equation}
which is the position of the $k^\text{th}$ agent rotating around a circle of radius $\rho_d$ and center at $c_d$ (see Eq.~\eqref{position}). This implies that, in the set $\Gamma_b$, all the agents converge to a common circle of radius $\rho_d = {\Omega}^{-1}_d > 0$ and center $c_d$. Moreover, by substituting \eqref{position_new} in \eqref{relation_control}, we get $\left(\partial W_1 / \partial \theta_k\right) = 0,~\forall k$, which defines the critical points of $W_1(\pmb{\theta})$ belonging to the set $\Gamma_b$. Since maximization of the potential $W_1(\pmb{\theta})$ corresponds to the global minimum of $U_1(\pmb{r}, \pmb{\theta}, \pmb{\omega})$, balanced formation is asymptotically stable (Lemma~\ref{lemma2}). This completes the proof.
\end{IEEEproof}

\begin{figure*}
\centerline{ \subfigure[]{\includegraphics[scale=0.3]{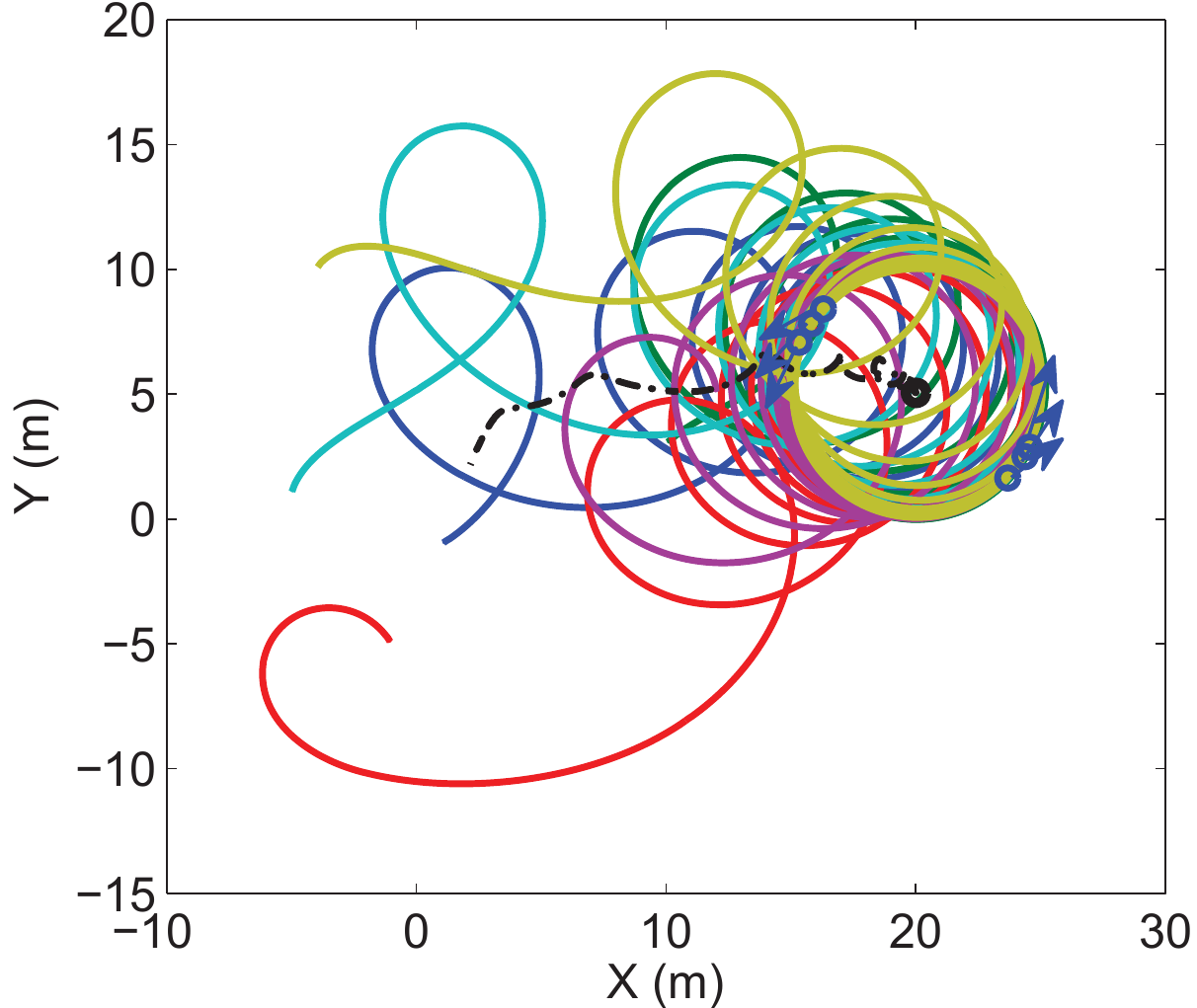}}\\
\subfigure[]{\includegraphics[scale=0.3]{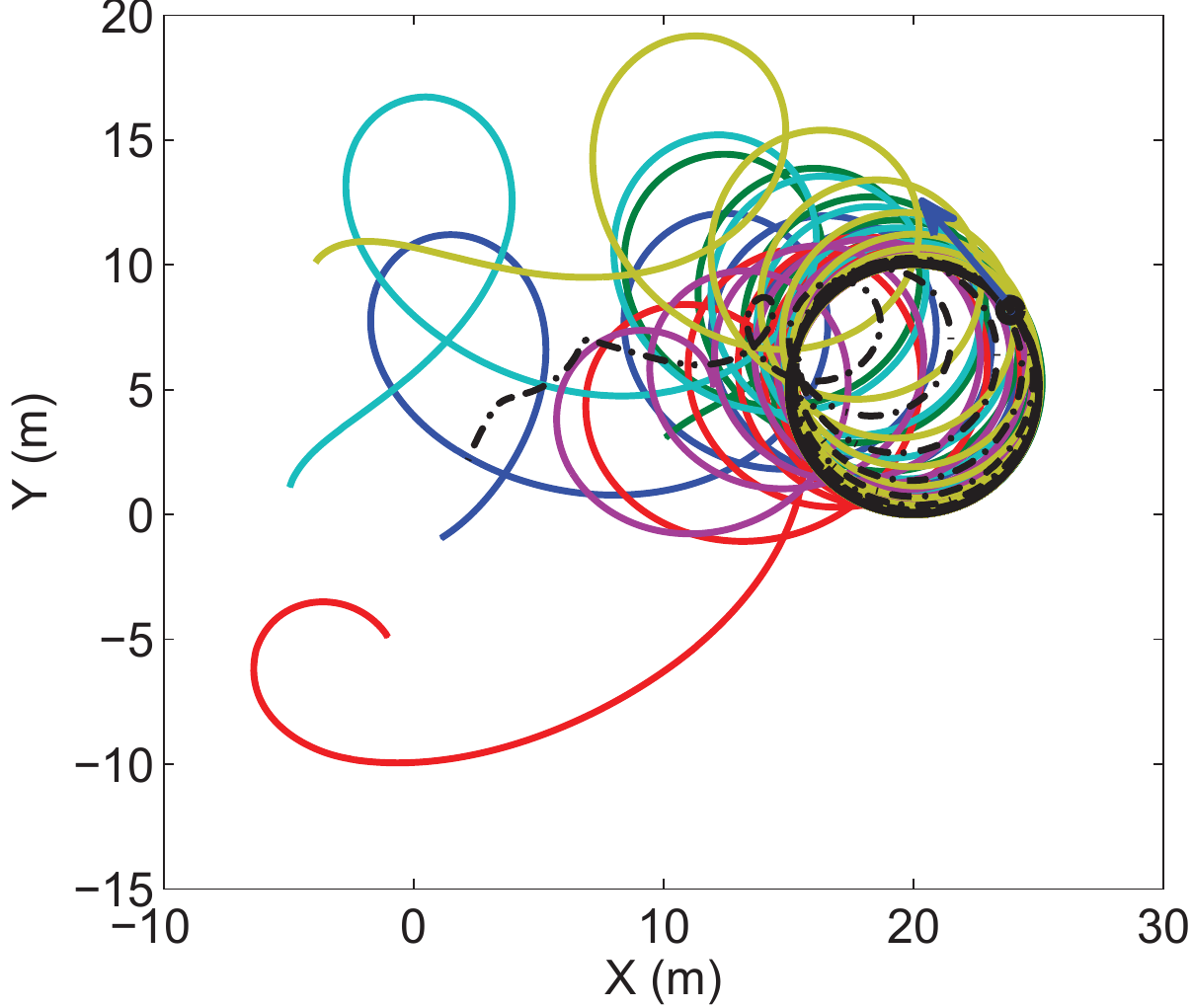}}\\
\subfigure[]{\includegraphics[scale=0.3]{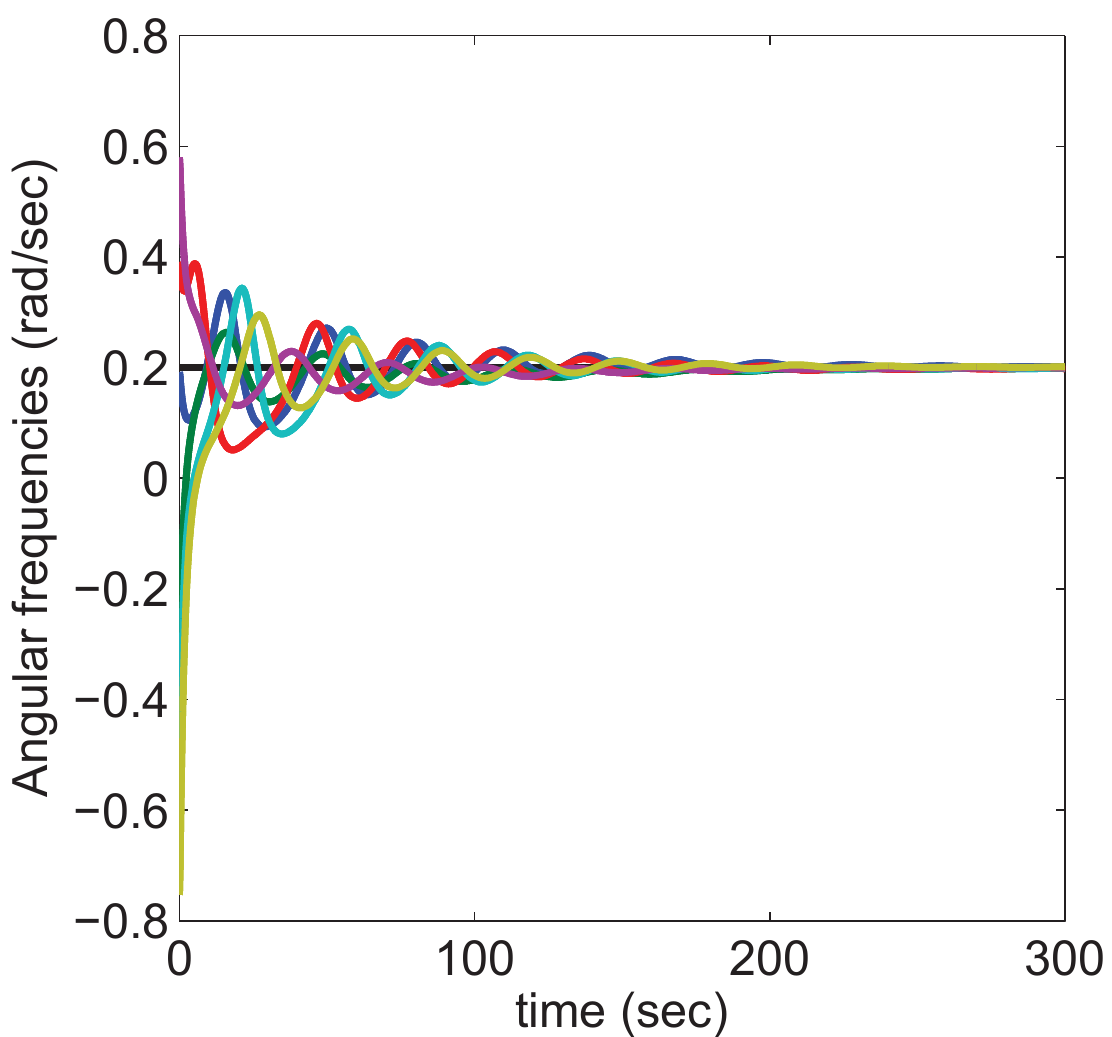}}}
\caption{Synchronization and balancing of $N = 6$ agents, connected by a graph as shown in Fig.1(b), around the common circle of desired radius $\rho_d = |\Omega_d|^{-1} = 5~ \text{m}$ and desired center $c_d = (20, 5)$. $(a)$ Balanced formation under the control law \eqref{control3} with $K = 0.5$ and $\kappa = 0.1$. $(b)$ Synchronized formation under the control law \eqref{control3} with $K = -1$ and $\kappa = 0.1$. $(c)$ Consensus of angular velocities at desired value $\Omega_d = 0.2~ \text{rad/sec}$ for balanced formation.}
\label{Stabilization on common circle}
\end{figure*}

\begin{thm}\label{Theorem4}
Let $\mathcal{L}$ be the Laplacian of an undirected and connected circulant graph $\mathcal{G} = (\mathcal{V}, \mathcal{E})$ with $N$ vertices. Consider system dynamics \eqref{modelNew} with control law \eqref{control3}. For $K < 0$ and $\kappa > 0$, all the agents converge to a circular formation in which they travel around a common circle of radius $\rho_d = \Omega_d^{-1} > 0$ and center $c_d$ in the anticlockwise direction with their phase angles in synchronized state.
\end{thm}

\begin{IEEEproof}
Consider a composite potential function
\begin{equation}
 U_2(\pmb{r}, \pmb{\theta}, \pmb{\omega}) = \kappa S(\pmb{r},\pmb{\theta}) - \rho_d K W_L(\pmb{\theta}) + {\rho}^3_d G(\pmb{\omega});~~K<0, \kappa > 0.\label{U_2}
\end{equation}
Under the control \eqref{control3}, the time derivative of $U_2(\pmb{r}, \pmb{\theta}, \pmb{\omega})$ along the trajectories of \eqref{modelNew}, yields
\begin{equation}
\dot{U}_2(\pmb{r}, \pmb{\theta}, \pmb{\omega}) = \dot{U}_1(\pmb{r}, \pmb{\theta}, \pmb{\omega}) = -\kappa {\rho}^4_d \sum_{k=1}^{N}
(\omega_k - \Omega_d)^2 \leq 0.
\end{equation}
Since $\dot{U}_2(\pmb{r}, \pmb{\theta}, \pmb{\omega}) = \dot{U}_1(\pmb{r}, \pmb{\theta}, \pmb{\omega})$, the proof follows the same steps as used to prove Theorem~\ref{Theorem3}. However, in this case, let $\Gamma_s$ be the largest invariant set in $\Gamma$ defined in \eqref{invariant_set_circular}. Thus, it can be concluded that all the agents converge to a common circle of radius $\rho_d = {\Omega}^{-1}_d > 0$ and center $c_d$ in the set $\Gamma_s$. Moreover, since minimization of the potential $W_1(\pmb{\theta})$ corresponds to the global minimum of $U_2(\pmb{r}, \pmb{\theta}, \pmb{\omega})$, synchronized formation is asymptotically stable (Lemma~\ref{lemma2}) in the set $\Gamma_s$. This completes the proof.
\end{IEEEproof}

{\it Example~2:} In this example, the simulation results are presented for the same $6$ agents as considered in Example~1.

Fig.~\ref{Stabilization on common circle} depicts the synchronization and balancing of the agents around a common circle at desired angular speed $\Omega_d = 0.2~ \text{rad/sec}$ (and hence desired radius $\rho_d = 5~\text{m}$) and desired center $c_d = (20, 5)$. Balanced formation is shown in Fig.~$5(a)$, and synchronized formation is shown in Fig.~$5(b)$. In Fig.~$5(c)$, the convergence of the angular speeds of the agents to a desired value $\Omega_d = 0.2~ \text{rad/sec}$, is shown in balanced formation only since the plot for synchronized formation is similar.

Note that these figures are obtained by setting different values of gains $K$ and $\kappa$. The selection of these gains is arbitrary and depends upon a particular problem of interest. By selecting gains $K$ and $\kappa$ appropriately, we can make the system of agents to stabilize in a desired formation, at faster or slower convergence rates.

It is worth noting that, in Fig.~ $5(a)$, the final position of the centroid of the group coincides with the center $c_d = (20, 5)$ of the common circle. This is due to the fact that, in balanced formation, the linear momentum $p_\theta = p_{1\theta} = 0$, which causes \eqref{position} to reduce to $c_d = (1/N)\sum_{k=1}^{N} r_k$ when summed over all $k$ on both the sides, which is the average position of all the agents, or the position of their centroid.

\begin{figure*}
\centerline{ \subfigure[]{\includegraphics[scale=0.3]{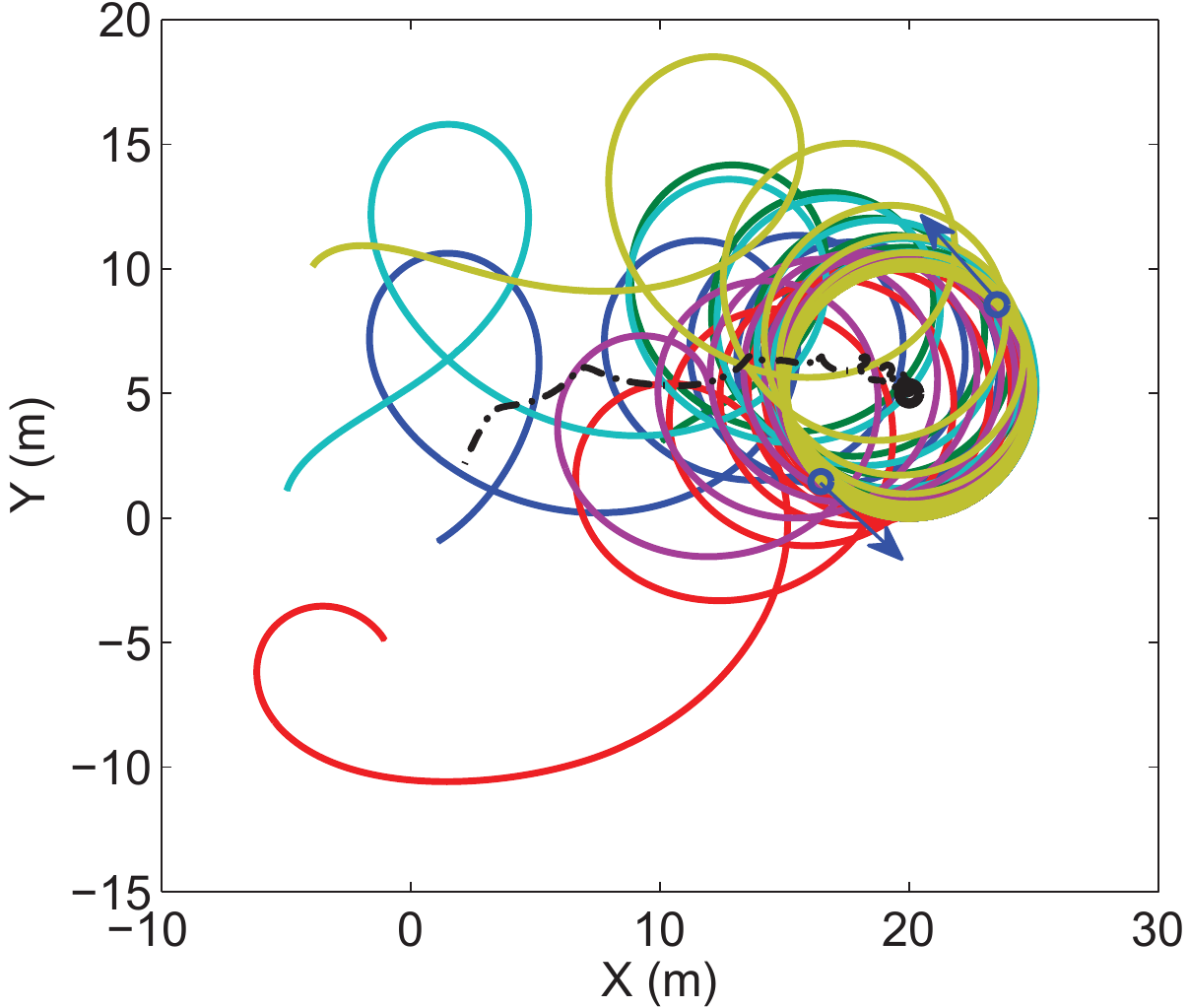}}\\
\subfigure[]{\includegraphics[scale=0.3]{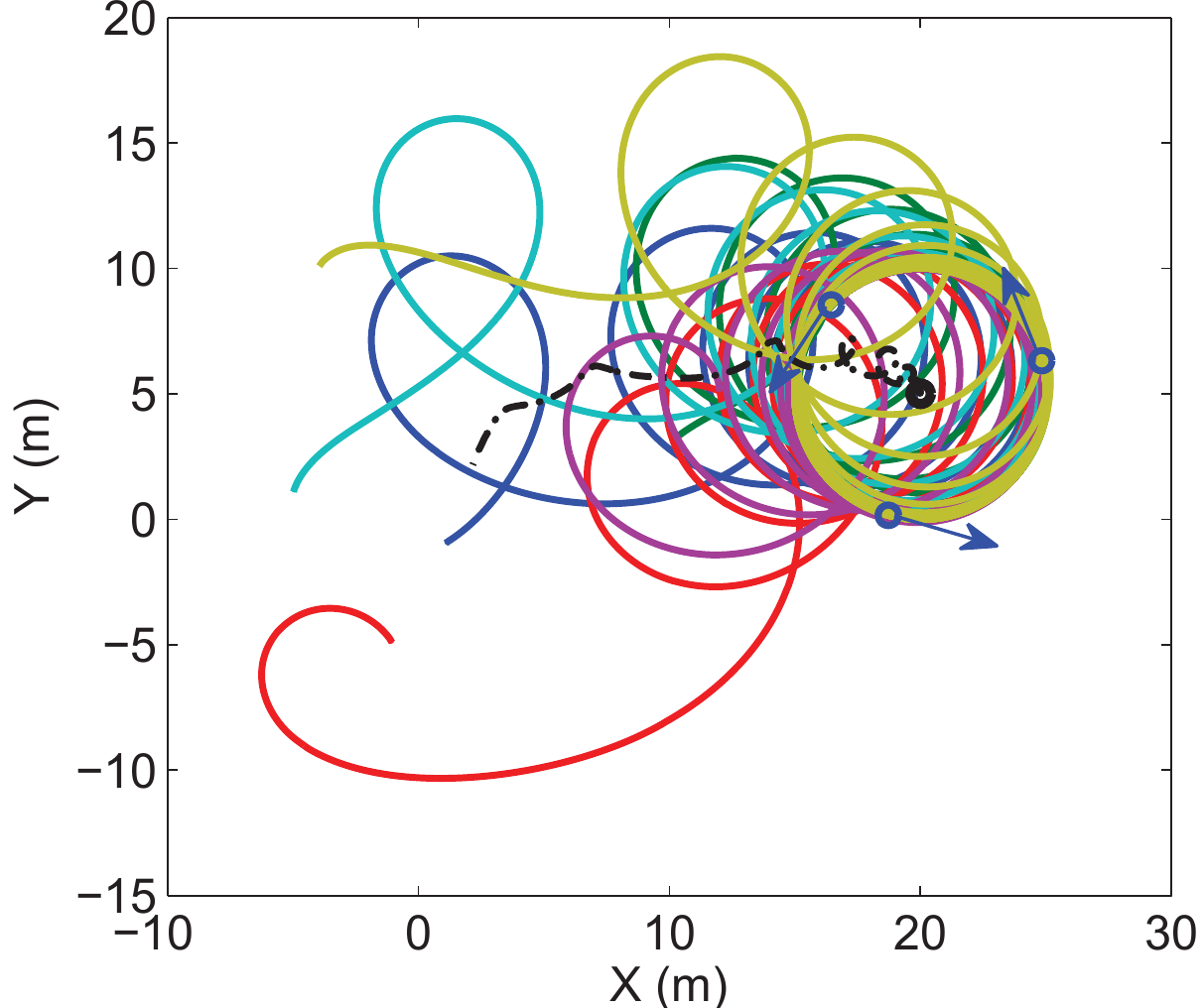}}\\
\subfigure[]{\includegraphics[scale=0.3]{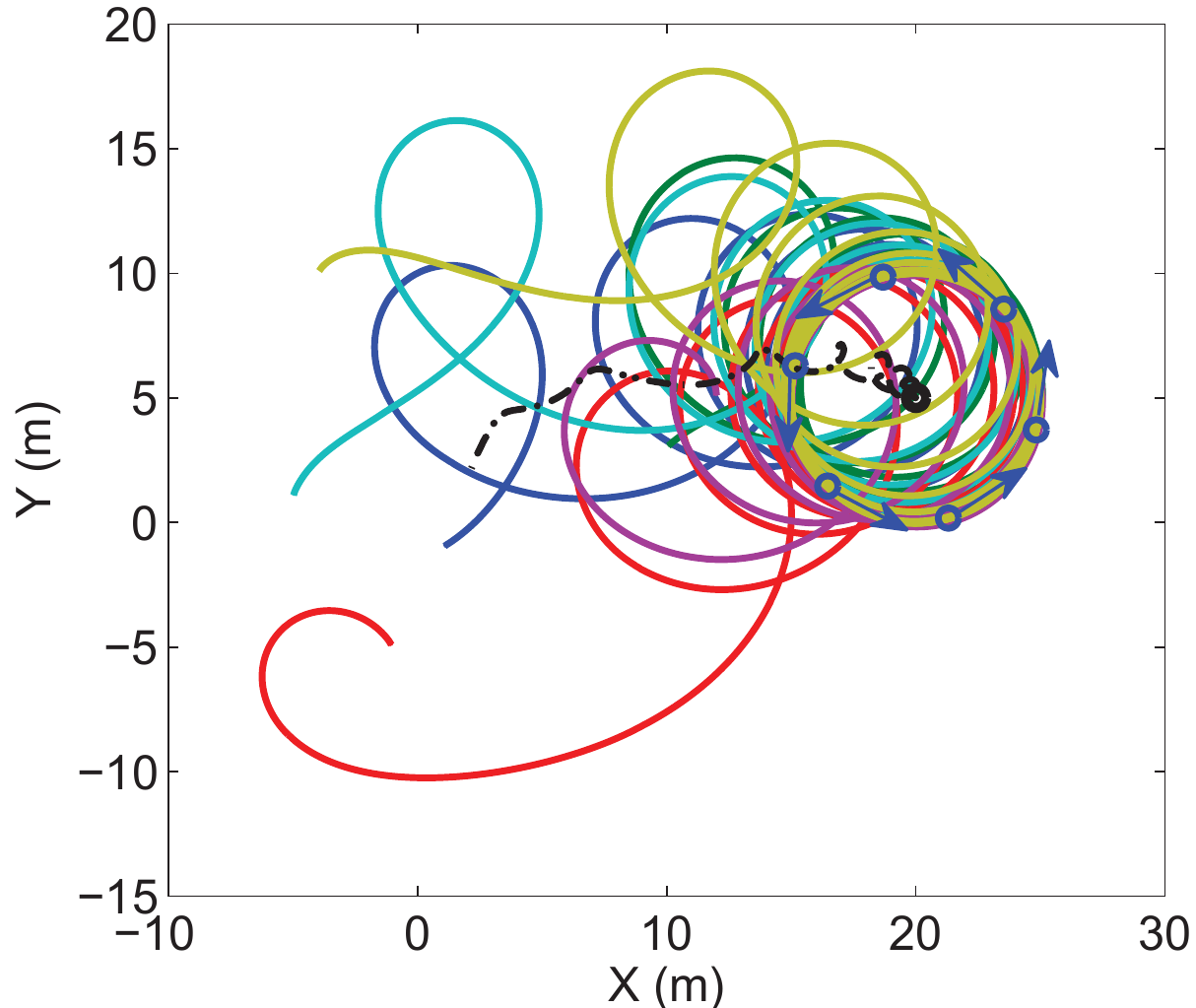}}}
\caption{Different symmetric balanced patterns of $6$ agents, connected by a graph as shown in Fig.1(b), around the common circle of desired radius $\rho_d = \Omega_d^{-1} = 5~ \text{m}$ and desired center $c_d = (20, 5)$, under the control law \eqref{control4} with $K=1, \kappa = K_m = 0.1$ for $1 \leq m \leq M-1$, and $K_M = -0.5$. $(a)$ $(2, 6)$ pattern. $(b)$ $(3, 6)$ pattern. $(c)$ $(6, 6)$ pattern: splay formation.}
\label{symmetric balanced patterns}
\end{figure*}

\section{Symmetric Balanced patterns Around the Common Circle}
In this  section, we aim to achieve symmetric balanced patterns of the agents around a common circle of desired radius $\rho_d = \omega^{-1}_d > 0$ as well as center $c_d$. These patterns are defined by the formations in which the agents are in phase balancing along with a symmetrical arrangement of their phase angles around the desired common circle \cite{Sepulchre2007}.

Let $1 \leq M \leq N$ be a divisor of $N$. A symmetric arrangement of $N$ phases consisting of $M$ clusters uniformly spaced around the common circle, each with $N/M$ synchronized phases, is called an $(M,N)$-pattern. For instance, the $(1,N)$-pattern corresponds to the synchronized state and the $(N,N)$-pattern corresponds to the so-called splay state, which is characterized by $N$ phases uniformly spaced around the common circle.

We now state the following lemma \cite{Paley2007} that is useful in proving the results of this section.

\begin{lem}\label{lemma3}
Let $1 \leq M \leq N$ be a divisor of $N$. An arrangement $\pmb{\theta}$ of $N$ phases is an $(M,N)$-pattern if and only if, for all $m \in \left\{1, \ldots, M-1\right\}$, the phase arrangement $m\pmb{\theta}$ is balanced and the phase arrangement $M\pmb{\theta}$ is synchronized.
\end{lem}


As mentioned in the above lemmas, the $m^\text{th}$ harmonic of the potential $W_m(\pmb{\theta})$ defined in \eqref{W_M} thus plays an important role in stabilizing symmetric phase patterns. Moreover, for various values of $m$, different symmetric balanced patterns arise as shown in Fig~ for $N=6$. See \cite{Sepulchre2007} for a more detailed description.

Following Lemma~\ref{lemma3}, we choose an $(M, N)$ phase potential as
\begin{equation}
W^{M,N}(\pmb{\theta}) = \sum_{m=1}^{M-1}\frac{K_m}{m^2} \left(\frac{{N}}{2}\lambda_\text{max} - W_m(\pmb{\theta})\right) - \frac{K_M}{M^2}W_M(\pmb{\theta}),\label{W_L_N}
\end{equation}
with $K_m > 0$ for $1 \leq m \leq M-1$, and $K_M < 0$. The global minimum of $W^{M,N}(\pmb{\theta})$ achieved (only) when $W_m(\pmb{\theta})$ is maximized for $1 \leq m \leq M-1$, and $W_M(\pmb{\theta})$ is minimized. This corresponds to the situation when $\pmb{\theta}, 2\pmb{\theta}, \ldots, (M-1)\pmb{\theta}$ are balanced and $M\pmb{\theta}$ is synchronized (Lemma~\ref{lemma2}), and hence the minimization of \eqref{W_L_N} gives rise to an $(M, N)$-pattern (Lemma~\ref{lemma3}).

Substituting for $W_m(\pmb{\theta})$ from \eqref{W_M} into \eqref{W_L_N}, yields
\begin{align}
\nonumber W^{M,N}(\pmb{\theta}) =& \frac{1}{2}\sum_{m=1}^{M-1}\frac{K_m}{m^2} \left({N}\lambda_\text{max} - \left<e^{im\pmb{\theta}}, \mathcal{L}e^{im\pmb{\theta}}\right>\right)\\
& -\frac{K_M}{2M^2}\left<e^{iM\pmb{\theta}}, \mathcal{L}e^{iM\pmb{\theta}}\right>,
\end{align}
whose time derivative along the dynamics \eqref{modelNew}, is given by
\begin{equation}
\label{W_L_M_N_dot}\dot{W}^{M,N}(\pmb{\theta}) = - \sum_{k=1}^{N}\sum_{m=1}^{M} K_m \left<ie^{im\theta_k}, \mathcal{L}_ke^{im\pmb{\theta}}\right>\omega_k,
\end{equation}
where, $\mathcal{L}_k$ is the $k^\text{th}$ row of the Laplacian $\mathcal{L}$.

\begin{thm}\label{Theorem5}
Let $\mathcal{L}$ be the Laplacian of an undirected and connected circulant graph $\mathcal{G} = (\mathcal{V}, \mathcal{E})$ with $N$ vertices. Consider system dynamics \eqref{modelNew} with control law
\begin{align}
\nonumber & u_k = -\kappa \rho_d(\omega_k - \Omega_d) \\
\label{control4} & + \Omega^2_d\left(\kappa \left<r_k-c_d, e^{i\theta_k}\right> + K \sum_{m=1}^{M} K_m \frac{\partial W_m}{\partial \theta_k}\right)
\end{align}
with $K_m > 0$ for $1 \leq m \leq M-1$, and $K_M < 0$. For $K>0$ and $\kappa > 0$, all the agents converge to a circular formation in which they travel around a common circle of radius $\rho_d = \Omega_d^{-1} > 0$ and center $c_d$ in the anticlockwise direction with their phase angles in the $(M,N)$-pattern.
\end{thm}

\begin{figure}[!t]
\centering
\includegraphics[scale=0.5]{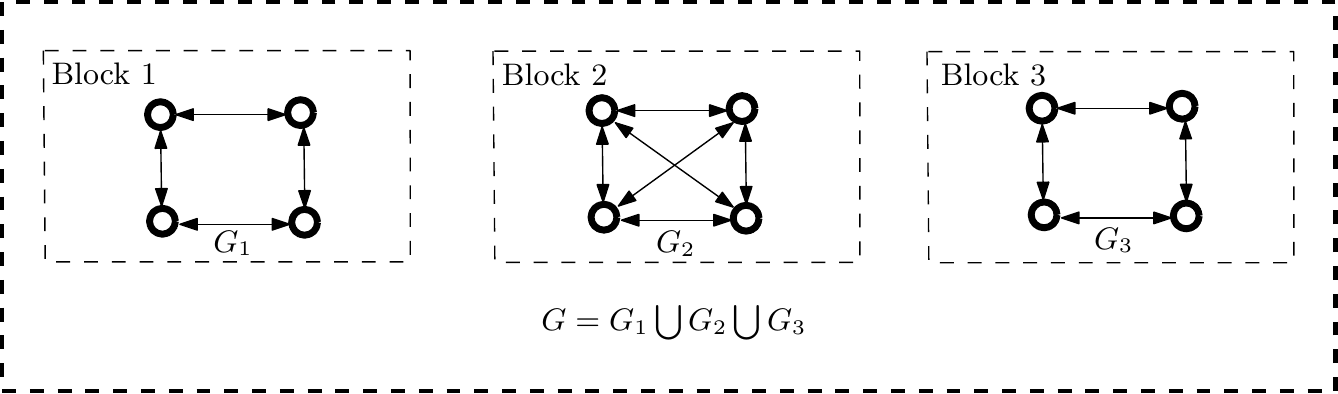}
\caption{Block interaction of $12$ agents with $3$ blocks, each containing $4$ agents.}
\label{Block interaction}
\end{figure}

\begin{IEEEproof}
Consider a composite potential function
\begin{equation}
\label{V}V(\pmb{r}, \pmb{\theta}, \pmb{\omega}) = \kappa S(\pmb{r}, \pmb{\theta}) + \rho_d K W^{M,N}(\pmb{\theta}) + {\rho}^3_d G(\pmb{\omega})
\end{equation}
with $K_m > 0$ for $1 \leq m \leq M-1$, and $K_M < 0$.
Using \eqref{Gdot}, \eqref{Sdot} and \eqref{W_L_M_N_dot}, the time derivative of the potential function $V(\pmb{r}, \pmb{\theta}, \pmb{\omega})$ along the dynamics \eqref{modelNew}, is
\begin{align}
\nonumber & \dot{V}(\pmb{r}, \pmb{\theta}, \pmb{\omega}) = \kappa
\sum_{k=1}^{N}\left<r_k-c_d, e^{i\theta_k}\right>\left(1 -
\rho_d\omega_k\right) \\
& + K \sum_{k=1}^{M} \sum_{m=1}^{N} K_m \left<ie^{im\theta_k}, \mathcal{L}_ke^{im\pmb{\theta}}\right>(-\rho_d\omega_k) + {\rho}^3_d \sum_{k=1}^{N}\left(\omega_k - \Omega_d\right) u_k.\label{Vdot1}
\end{align}
Note that
\begin{equation}
\label{relation_new}\sum_{k=1}^{N}\left<ie^{im\theta_k}, L_ke^{im\pmb{\theta}}\right> = - \sum_{k=1}^{N}\sum_{j \in \mathcal{N}_k} \sin \left(m(\theta_j - \theta_k)\right) = 0
\end{equation}
Using \eqref{relation_new}, \eqref{Vdot1} can be rewritten as
\begin{align}
\nonumber\dot{V}(\pmb{r}, \pmb{\theta}, \pmb{\omega}) &= \sum_{k=1}^{N}\left(\kappa \left<r_k-c_d, e^{i\theta_k}\right> + K \sum_{m=1}^{M} K_m \left<ie^{im\theta_k}, \mathcal{L}_ke^{im\pmb{\theta}}\right>\right)\\
& \times \left(1 - \rho_d\omega_k \right) + {\rho}^3_d \sum_{k=1}^{N}\left(\omega_k - \Omega_d\right) u_k.
\end{align}
Under the control law \eqref{control4}, the time derivative of $V(\pmb{r}, \pmb{\theta}, \pmb{\omega})$ results in
\begin{equation}
\dot{V}(\pmb{r}, \pmb{\theta}, \pmb{\omega}) = \dot{U}_1(\pmb{r}, \pmb{\theta}, \pmb{\omega}) = - \kappa {\rho}^4_d \sum_{k=1}^{N} (\omega_k - \Omega_d)^2 \leq 0.
\end{equation}
Since $\dot{V}(\pmb{r}, \pmb{\theta}, \pmb{\omega}) = \dot{U}_1(\pmb{r}, \pmb{\theta}, \pmb{\omega})$, the proof follows the same steps as used to prove Theorem~$3$. However, in this case, let $\Xi$ be the largest invariant set in $\Gamma$ defined in \ref{invariant_set_circular}. Thus, it can be concluded that all the agents converge to a common circle of radius $\rho_d = \Omega_d > 0$ and center $c_d$ in the set $\Xi$. Moreover, it follows from Lemma~\ref{lemma2} that the maximization of $W_m(\theta)$ for $1 \leq m \leq M-1$, and minimization of $W_M(\theta)$ corresponds to the global minimum of the potential $V(\pmb{r}, \pmb{\theta}, \pmb{\omega})$. Thus, the phase arrangements $\pmb{\theta}, 2\pmb{\theta}, \ldots, (M-1)\pmb{\theta}$ are balanced and $M\pmb{\theta}$ is synchronized in the set $\Xi$, and hence give rise to an $(M, N)$ phase pattern (Lemma~\ref{lemma3}). This completes the proof.
\end{IEEEproof}

{\it Example~3:} In this example, the simulation results are presented for the $6$ agents considered in Example~1. Fig.~\ref{symmetric balanced patterns} shows the different symmetric balanced patterns of the agents around the common circle of desired radius $\rho_d = \Omega_d^{-1} = 5~ \text{m}$ and center $c_d = (20, 5)$. The arrangement in Fig.~$6(c)$ is the splay state, in which the $6$ agents are at equal angular separation of $60^\circ$, as expected.

\begin{remark}
We may assume that agents move on different altitudes as in \cite{Seyboth2014} when they are in synchronization around a common circular orbit. Approaches to collision avoidance among agents will be explored in future.
\end{remark}

\section{Achieving Coordinated Subgroups}
Motivated by mobile sensor network applications, as discussed in \cite{Sepulchre2008}, in this section, we propose control law that uses multiple graphs at a time to yield multi-level sensing patterns in a group of agents. By using such a multi-level interaction among agents, a formation in which the agents are arranged in subgroups around different circular orbits with a symmetric pattern of their phase angles, can be obtained. In this context, the different graphs, by considering both intra- and inter subgroup coordination, are used in the steering control law, so that subgroups can be formed and regulated. A particular application of such formations can be found in the optimal mobile sensor coverage problem where it is required for the sensors to move around closed curves with coordinated phasing for maximizing information in data collected in a spatially and temporally varying field (for instance, from an ocean) \cite{Leonard2007}.

In order to use multi-level sensing networks, we divide the group of $N$ agents into $B$ blocks (subgroups) and refer to each block by its block index $1 \leq b \leq B$. For the sake of convenience, we assume that there is no interaction between subgroups unless specifically mentioned. Let the graph $\mathcal{G}_b$ describe the interaction between all the agents in block $b$. Collectively, the set of all interaction is defined by the graph $G\triangleq \bigcup_{b=1}^{B} \mathcal{G}_b$. In this situation, the Laplacian $\hat{\mathcal{L}}$ corresponding to graph $\mathcal{G}$ is a block-diagonal matrix, each block of which represents the Laplacian of the graph $\mathcal{G}_b$. For instance, consider a group of $12$ agents divided into 3 blocks containing 4 agents in each block. Let their interaction network be represented by a graph $\mathcal{G} = \mathcal{G}_1 \bigcup \mathcal{G}_2 \bigcup \mathcal{G}_3$ as shown in Fig.~\ref{Block interaction}. The Laplacian matrix corresponding to this block interaction is given by
\begin{equation}
\label{Laplacian_coordinated}\hat{\mathcal{L}} = \left[
\begin{array}{r|r|r}
     \mathcal{L}_{\mathcal{G}_1} & 0_{4\times4} &  0_{4\times4}\\ \hline
0_{4\times4} &   \mathcal{L}_{\mathcal{G}_2}    &  0_{4\times4}\\ \hline
0_{4\times4} & 0_{4\times4} &   \mathcal{L}_{\mathcal{G}_3}
\end{array}
\right]
\end{equation}
where, $0_{4\times4}$ represents a ${4\times4}$ zero matrix, and $\mathcal{L}_{\mathcal{G}_1} = \mathcal{L}_{\mathcal{G}_3} = \text{circ}(2, -1,  0, -1)$ and $\mathcal{L}_{\mathcal{G}_2} = \text{circ}(3, -1,  -1,  -1)$  are the Laplacian of the subgraphs $\mathcal{G}_1 (= \mathcal{G}_3)$ and $\mathcal{G}_2$, respectively. Since $\hat{\mathcal{L}}$ is a block diagonal matrix, its eigenvalues are the union of the eigenvalues of all block diagonal matrices \cite{Rorres2011}. In this illustration, although we assume that each subgroup is of the same size, however, this is not required.

We further assume that each agent is assigned to one and only one block, so that $\sum_{b=1}^{B} N_b = N$, where $N_b$ is the number of agents in the $b^\text{th}$ block. Also, let $F^b = \left\{f^b_1, \ldots, f^b_{N^b}\right\}$ be the set of agent indices, and $\rho^b_d = \left\{\Omega^b_d\right\}^{-1} > 0$ and $c^b_d$, respectively, be the desired radius, and center of the common circular orbit associated with the $b^\text{th}$ block.

Based on these notations, we now consider the following two cases of coordinated subgroups with respect to the desired radius $\rho^b_d$ for a reason which will be apparent little later.

\subsection{Case~1: $\rho^b_d$ is different for all $1 \leq b \leq B$}
In this situation, we have the following corollaries to Theorems~\ref{Theorem3}, \ref{Theorem4} and \ref{Theorem5}.

\begin{cor}\label{cor1}
For $k \in F^b$, consider system dynamics \eqref{modelNew} with control law
{\small \begin{equation}
u_k = -\kappa \rho^b_d(\omega_k - \Omega^b_d) + \left\{\Omega^b_d\right\}^2 \left(\kappa \left<r_k-c^b_d, e^{i\theta_k}\right> + K \left<ie^{i\theta_k}, \hat{\mathcal{L}}_ke^{i\pmb{\theta}}\right>\right),\label{control5}
\end{equation}}
\noindent where, $\hat{\mathcal{L}}_k$ is the $k^\text{th}$ row of the Laplacian $\hat{\mathcal{L}}$ of a graph $\mathcal{G} = \bigcup_{b=1}^{B} \mathcal{G}_b$ with $\mathcal{G}_b$ being an undirected and connected circulant subgraph. For $K>0$ and $\kappa > 0$, all the agents belonging to the $b^\text{th}$ block converge to a circular formation in which they move on a circle of radius $\rho^b_d = \left\{\Omega^b_d\right\}^{-1} > 0$ and center $c^b_d$ in the anticlockwise direction with their phase angles in the balanced state.
\end{cor}

\begin{IEEEproof}
Consider a composite potential function
{\small \begin{align}
\nonumber U^b_1(\pmb{r}, \pmb{\theta}, \pmb{\omega}) =& \kappa S^b(\pmb{r},\pmb{\theta}) + \rho^b_d K \left(\frac{N}{2}\lambda_\text{max} - \hat{W}_1(\pmb{\theta})\right)\\
& + \left\{\rho^b_d\right\}^3 G^b(\pmb{\omega});~~~ \kappa > 0, K > 0,\label{U^b_1}
\end{align}}
where,
{\small
\begin{align}
\nonumber S^b(\pmb{r},\pmb{\theta}) &= \frac{1}{2}\sum_{k=1}^{N} \left<r_k-c^b_d + i\rho^b_0e^{i\theta_k}, r_k-c^b_d + i\rho^b_de^{i\theta_k}\right>,\\
\nonumber \hat{W}_1(\pmb{\theta})   &=  (1/2)\left<e^{i\pmb{\theta}}, \hat{\mathcal{L}}e^{i\pmb{\theta}}\right>,\\
\nonumber G^b(\pmb{\omega}) &= \frac{1}{2}\sum_{k=1}^{N}\left(\omega_k - \Omega^b_d\right)^2,
\end{align}
} \noindent and $\lambda_\text{max}$ is the maximum eigenvalue of $\hat{\mathcal{L}}$. Since \eqref{U^b_1} has a structure similar to \eqref{U_1}, the proof is similar to the proof of Theorem~\ref{Theorem3} and hence omitted.
\end{IEEEproof}

Similarly the corollaries to Theorems~\ref{Theorem4} and \ref{Theorem5} can be stated, and are not described here to avoid repetition.

\begin{figure*}
\centerline{ \subfigure[]{\includegraphics[scale=0.3]{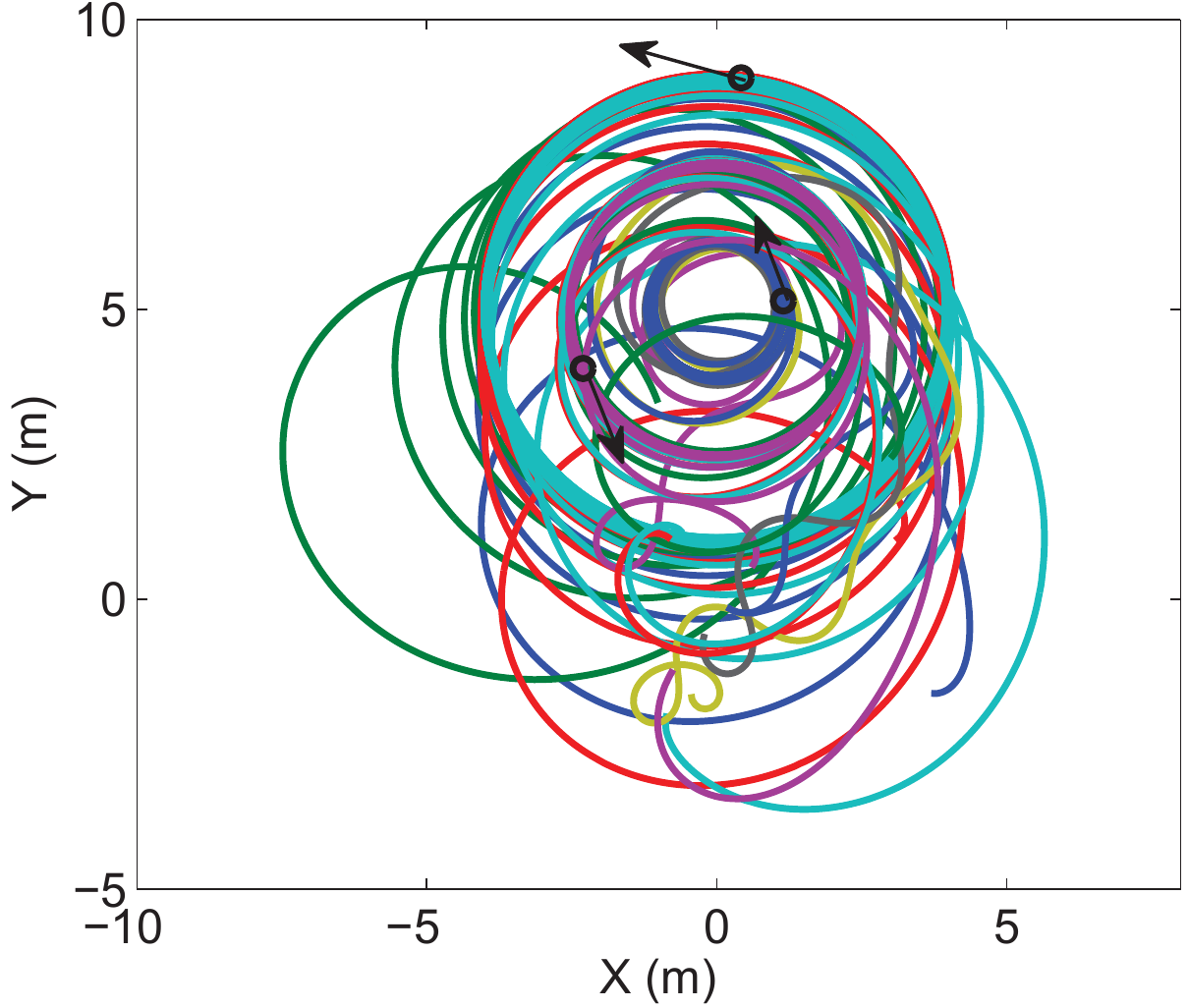}}\\
\subfigure[]{\includegraphics[scale=0.3]{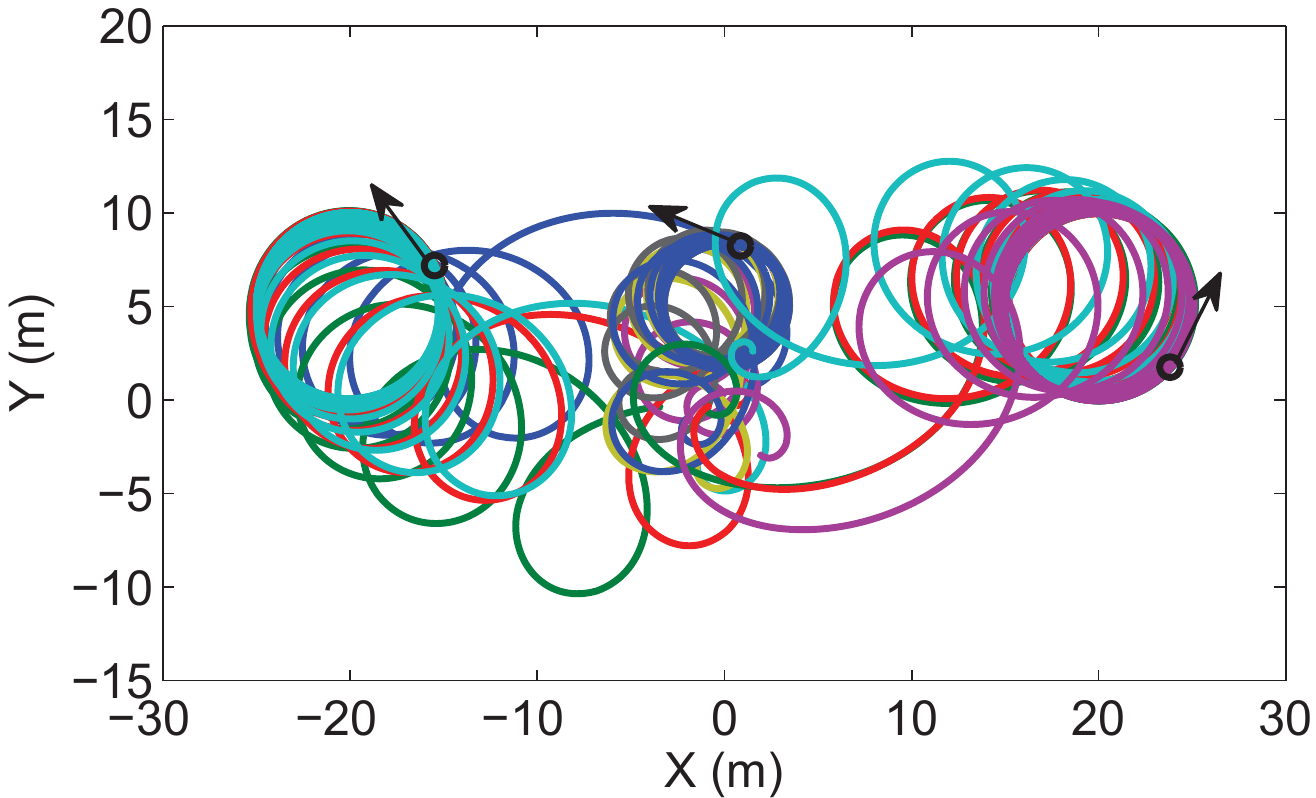}}\\
\subfigure[]{\includegraphics[scale=0.3]{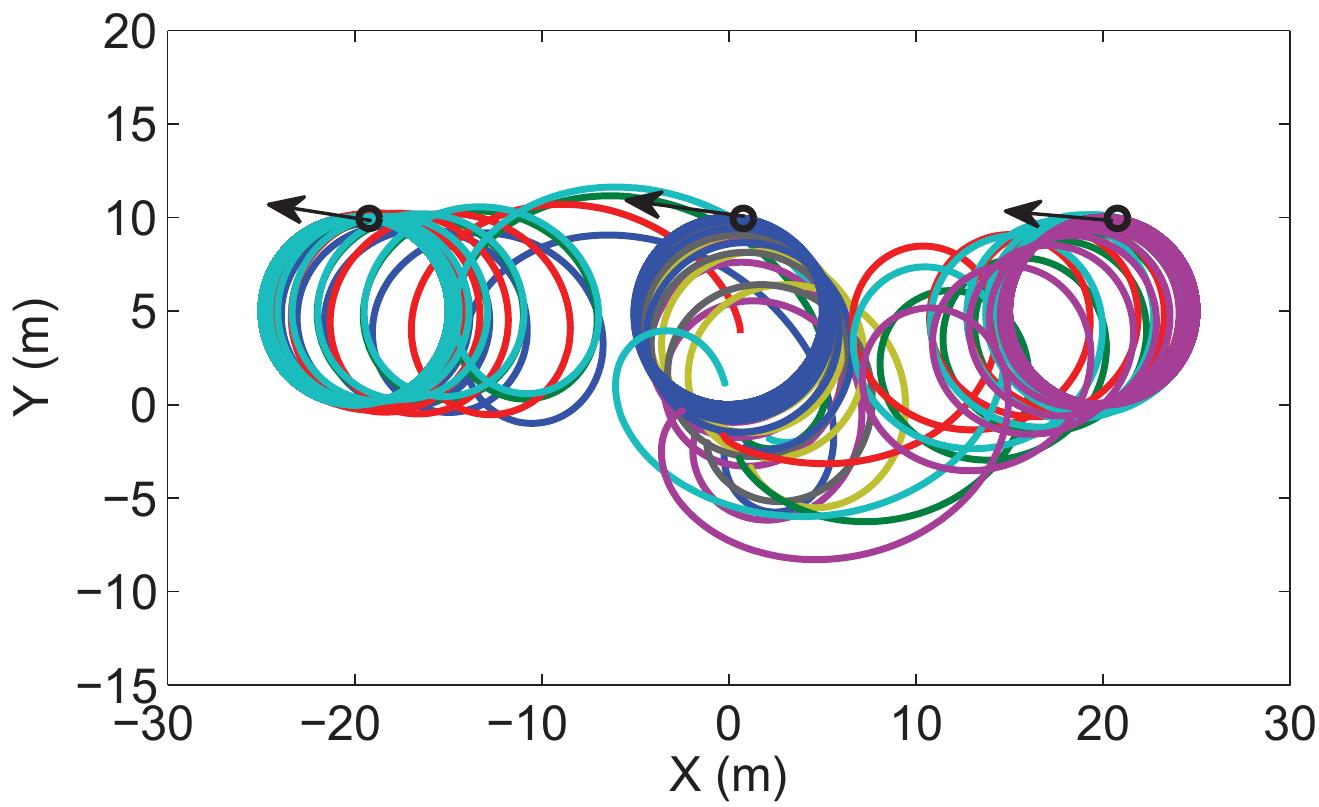}}}
\caption{Synchronization of $12$ agents in 3 different groups, with 4 agents in each group, under the control laws \eqref{control5} and \eqref{control7} with $K=-1$ and $\kappa=0.5$. $(a)$ Different desired radius and the same desired center when the agents interact only within subgroups. $(b)$ Different desired radius and centers when the agents interact only within subgroups. $(c)$ Same desired radius and different desired centers when the agents interact not only within subgroups but also among subgroups.}
\label{Stabilization of synchronized formation in groups}
\end{figure*}

\begin{figure*}
\centerline{\subfigure[]{\includegraphics[scale=0.3]{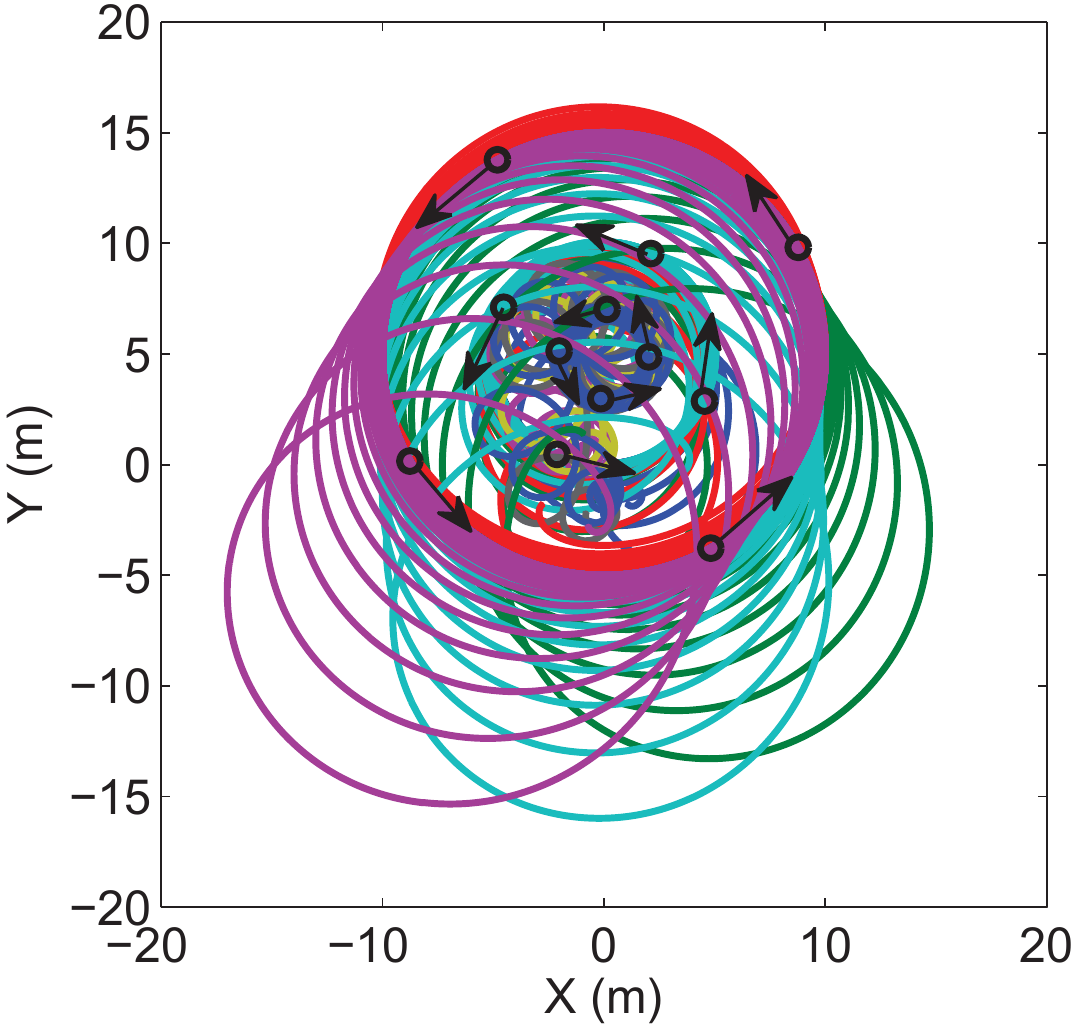}}\\
\subfigure[]{\includegraphics[scale=0.3]{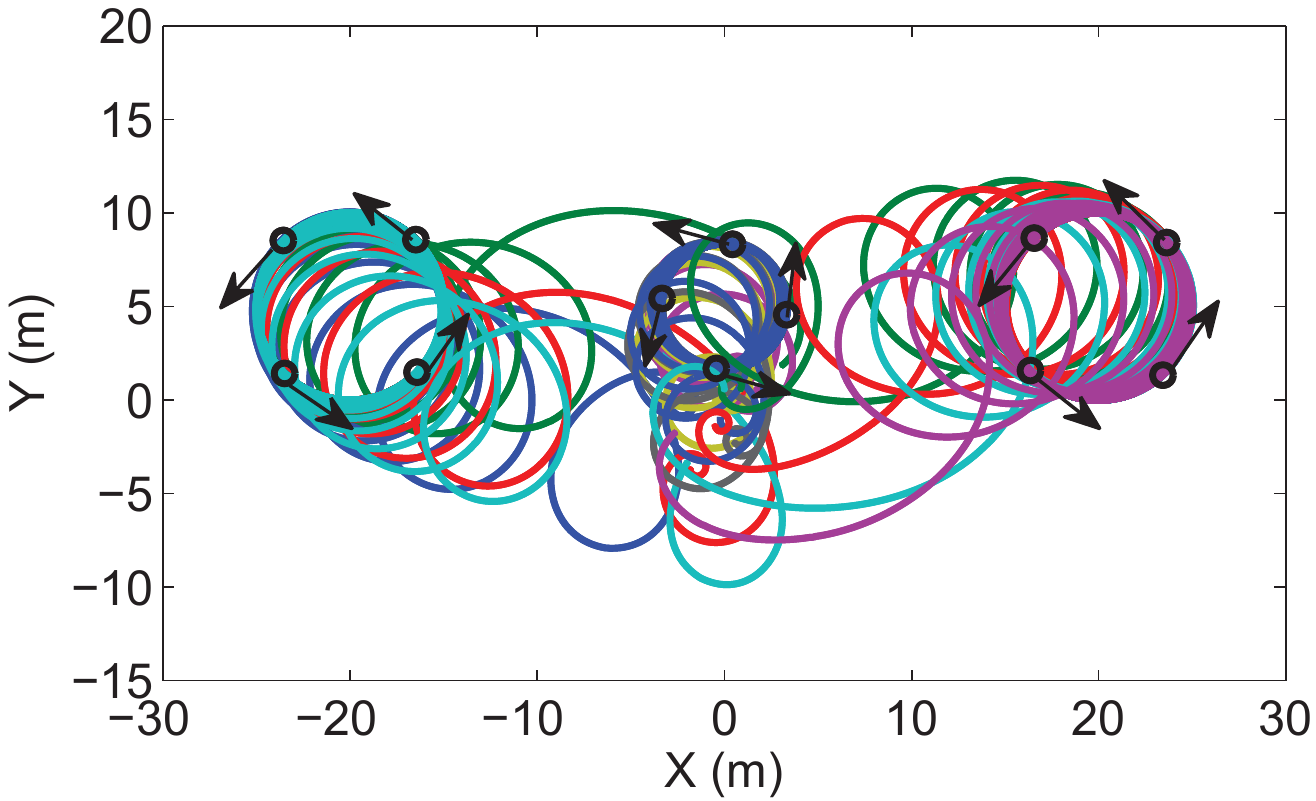}}\\
\subfigure[]{\includegraphics[scale=0.3]{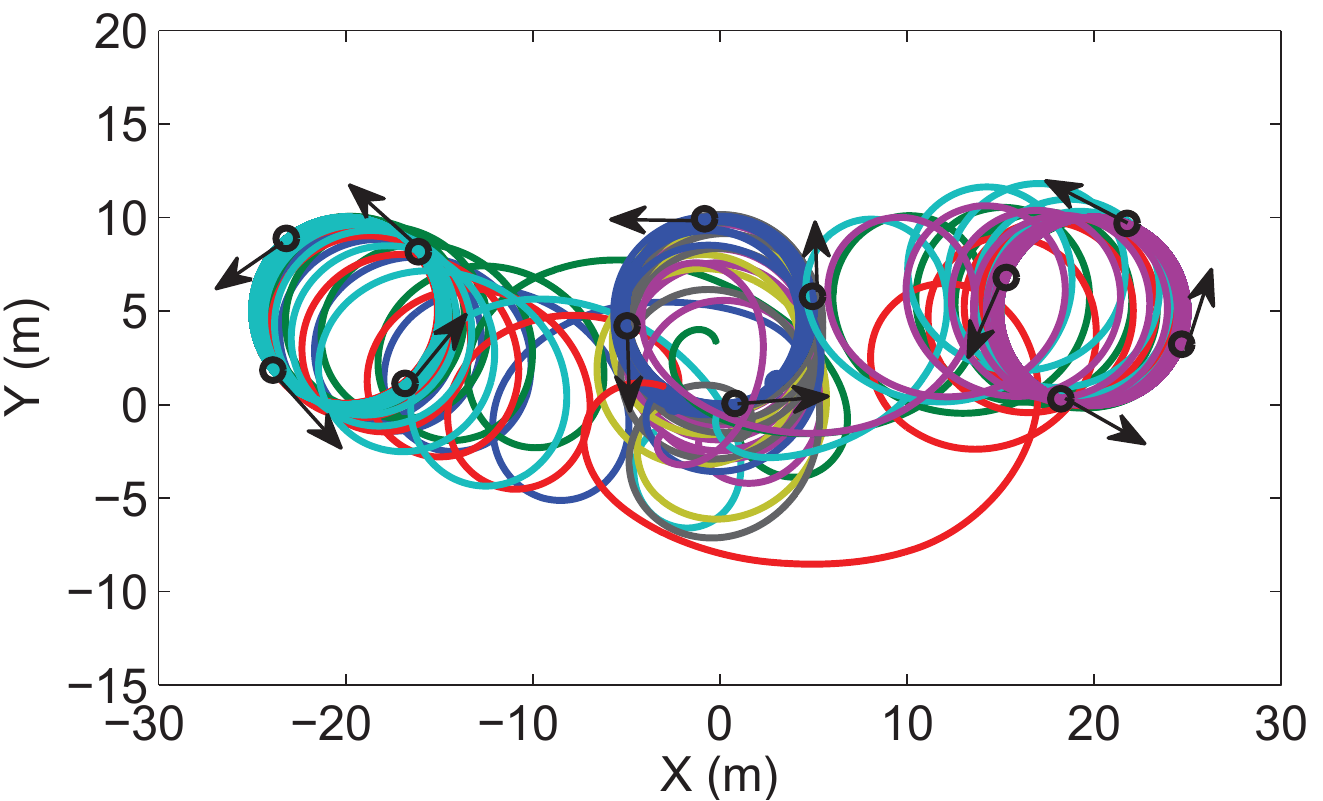}}}
\caption{Splay formation of $12$ agents in 3 different groups, with 4 agents in each group. $(a)$ Different desired radius and the same desired center when the agents interact only within subgroups. $(b)$ Different desired radius and centers when the agents interact only within subgroups. $(c)$ Same desired radius and centers when the agents interact not only within subgroups but also among subgroups.}
\label{Stabilization of splay formation in groups}
\end{figure*}

\subsection{Case~2: $\rho^b_d = \rho_d$ for all $1 \leq b \leq B$}
In such a situation, by assuming that the agents within a block can also interact with the agents in other blocks in addition to their inter block interaction, it is possible to use multi-level sensing network in a single control algorithm since the phase potentials corresponding to the different communication topologies can be combined to get a single phase potential function. For instance, consider a group of 12 agents as shown in Fig.~\ref{Block interaction}, now with two types of sensing networks comprising an intra- and an inter subgroup coordination. Assume that the intra subgroup interaction is all-to-all and the inter subgroup interaction is as shown Fig.~\ref{Block interaction}. Let $\tilde{\mathcal{L}} = NP (= NI_N - \pmb{1}\pmb{1}^T)$, and $\hat{\mathcal{L}}$, given by \eqref{Laplacian_coordinated}, be the Laplacian corresponding to the intra- and inter subgroup coordination, respectively. Then, with such a multi-level interaction, the potential function in \eqref{U^b_1} becomes
\begin{equation}
\overline{U}^b_1(\pmb{r}, \pmb{\theta}, \pmb{\omega}) = \kappa S^b(\pmb{r},\pmb{\theta}) + \rho_d K \left(\frac{N}{2}\lambda_\text{max} - \overline{W}_1(\pmb{\theta})\right) + \rho^3_d\ G(\pmb{\omega}),\label{U_hat_2_b}
\end{equation}
where,
\begin{equation}
\overline{W}_1(\pmb{\theta}) =  \frac{1}{2}\left<e^{i\pmb{\theta}}, (\hat{\mathcal{L}} + \tilde{\mathcal{L}})e^{i\pmb{\theta}}\right>.
\end{equation}
The potential $\overline{U}^b_1(\pmb{r}, \pmb{\theta}, \pmb{\omega})$ is minimized by choosing the control
\begin{align}
\nonumber u_k  = & -\kappa \rho_d(\omega_k - \Omega_d) + \left\{\Omega_d\right\}^2 \\
 \times & \left[\kappa \left<r_k-c^b_d, e^{i\theta_k}\right> + K \left(\left<ie^{i\theta_k}, \left(\hat{\mathcal{L}}_k + \tilde{\mathcal{L}}_k\right)e^{i\pmb{\theta}}\right>\right)\right],\label{control7}
\end{align}
which accounts for both intra- and inter subgroups coordination. Note that, unlike previous case of the different radius of the circular orbits, here we can combine all the phase stabilizing potentials independently to get a single potential function since $\rho^b_d = \rho_d$ for all $b =1, \ldots, B$. Thus, under the control \eqref{control7}, the phase arrangement in which the entire group, as well as each block, are in balanced formation, is obtained. The same behavior of agent's phase arrangement in synchronized and splay formations can be depicted under suitable control laws.


\begin{remark}
It is evident from the above discussion that the second order rotational model is adequate to control the phase arrangement as well as the radius of the desired circular orbit. Thus, unlike \cite{Sepulchre2008}, the class of collective circular motion studied in this paper shows an interesting possibilities for the unmanned vehicles to expand and contract their formations about a desired location so as to better explore the search area, and thus more desirable for the applications to mobile sensor networks.
\end{remark}

{\it Example~5:} In this example, simulation results are presented for the above considered $N =12$ agents whose initial positions, initial heading angles and initial angular velocities are randomly generated.

At first, we assume that the agents interact only in blocks according to Fig.~\ref{Block interaction}. In such a situation, synchronized and splay formations of the agents in three groups are shown in Figs.~\ref{Stabilization of synchronized formation in groups}, and \ref{Stabilization of splay formation in groups}, respectively. The synchronized and splay formations of the agents, on the circles of different desired radius and the same desired center, are shown in Figs.~$8(a)$ and $9(a)$, respectively, while the same, on the circles of different desired radius as well as centers is shown in Figs.~$8(b)$ and $9(b)$, respectively. Since this case corresponds to different $\rho^d_d$, the agents are in synchronized and in splay formations within subgroups only.

Next, we assume that the agents interact within subgroups according to Fig.~\ref{Block interaction} as well as among subgroups according to an all-to-all communication topology as discussed above. In this situation, the synchronized and splay formations of the agents, on the circles of the same desired radius and different desired centers, are shown in Figs.~$8(c)$ and $9(c)$, respectively. Since the radius of all the circles  corresponding to each group is same, the agents are collectively in synchronized and splay formations. In other words, if the heading phasors of all the agents were plotted on the same circle, then the resulting pattern would be synchronized or in splay state.


\section{Conclusions}
A Laplacian-based control design methodology to stabilize synchronized and balanced collective circular motions of a group of agents with second-order rotational dynamics under limited communication topology, which is represented by a time-invariant undirected graph, has been proposed in this paper. In particular, the collective motion of agents around different circles or around a common circle at a desired angular velocity have been investigated. The second-order feedback control laws have been derived from composite potential functions, which reach their minimum in the desired configuration of the agents. It has been shown that the second-order feedback controls are adequate to regulate the orientations as well as the angular velocities of the agents. The LaSalle's invariance principle has been used extensively to prove the asymptotic stability of the desired circular formation under the proposed control scheme. Moreover, the use of multi-level interaction networks to obtain various symmetric circular formations suitable for applications to mobile sensors has been discussed in the context of better exploration of a search area. A piece of work in future to pursue time-varying communication topology and the collision avoidance.



\end{document}